\documentclass[nonacm]{acmart}

\usepackage{amsmath}
\usepackage{url}
\usepackage{multirow}
\usepackage[all]{nowidow}
\usepackage{algorithm}
\usepackage{amsfonts}
\usepackage[noend]{algpseudocode}
\usepackage{cleveref}
\usepackage{listings}
\usepackage{subcaption}
\usepackage{tikz}
\usepackage{xcolor}
\usepackage{comment}
\usepackage{xspace}

\usetikzlibrary{shapes,arrows}
\newcommand{\toolname}{\textit{Beryllium}\xspace}

\definecolor{main-color}{rgb}{0.6627, 0.7176, 0.7764}
\definecolor{back-color}{rgb}{0.1686, 0.1686, 0.1686}
\definecolor{string-color}{rgb}{0.9333, 0.5254, 0.345}
\definecolor{key-color}{rgb}{0.8, 0.47, 0.196}

\lstdefinestyle{mystyle}
{
    basicstyle = {\footnotesize \color{back-color}},
    literate={\ \ }{{\ }}1,
    tabsize=3,
    stringstyle = {\color{string-color}},
    keywordstyle = {\color{key-color}},
    keywordstyle = [2]{\color{blue}},
    keywordstyle = [3]{\color{yellow}},
    keywordstyle = [4]{\color{teal}},
    otherkeywords = {;,<<,>>,++,==,=,-,+, .,0,1,2,3,4,5,6,7,8,9},
    morekeywords = [2]{;},
    morekeywords = [3]{<<, >>},
    morekeywords = [4]{++, 0,1,2,3,4,5,6,7,8,9},
}

\lstdefinestyle{mystyle-java}
{
    language = JAVA,
    basicstyle = {\footnotesize \color{back-color}},
    literate={\ }{{\ }}1,
    tabsize=3,
    stringstyle = {\color{string-color}},
    keywordstyle = {\color{key-color}},
    keywordstyle = [2]{\color{blue}},
    keywordstyle = [3]{\color{yellow}},
    keywordstyle = [4]{\color{teal}},
    otherkeywords = {;,<<,>>,++,==,=,-,+, ., 0,1,2,3,4,5,6,7,8,9},
    morekeywords = [2]{;},
    morekeywords = [3]{<<, >>},
    morekeywords = [4]{0,1,2,3,4,5,6,7,8,9},
}

\AtBeginDocument{%
  \providecommand\BibTeX{{%
    \normalfont B\kern-0.5em{\scshape i\kern-0.25em b}\kern-0.8em\TeX}}}

\setcopyright{acmcopyright}
\copyrightyear{2023}
\acmYear{2023}
\acmDOI{XXXXXXX.XXXXXXX}

\acmConference[ACM Transactions on Software Engineering and Methodology]{Make sure to enter the correct
  conference title from your rights confirmation emai}{2023}{Victoria, AU}
\acmPrice{15.00}
\acmISBN{978-1-4503-XXXX-X/18/06}
\begin{document}

\title{Beryllium: Neural Search for Algorithm Implementations}

\author{Adithya Kulkarni}
\authornote{The first three authors contributed equally to this research.}
\email{aditkulk@iastate.edu}
\author{Mohna Chakraborty}
\authornotemark[1]
\email{mohnac@iastate.edu}
\author{Yonas Sium}
\authornotemark[1]
\email{yas@iastate.edu}
\affiliation{
  \institution{Iowa State University}
  % \streetaddress{P.O. Box 1212}
  \city{Ames}
  \state{Iowa}
  \country{USA}
  % \postcode{50011}
}

\author{Sai Charishma Valluri}
\affiliation{
  \institution{Iowa State University}
  \city{Ames}
  \country{USA}
\email{svalluri@iastate.edu}}

\author{Wei Le}
\affiliation{
  \institution{Iowa State University}
  \city{Ames}
  \country{USA}
\email{weile@iastate.edu}}

\author{Qi Li}
\affiliation{
  \institution{Iowa State University}
  \city{Ames}
  \country{USA}
\email{Qli@iastate.edu}}

\renewcommand{\shortauthors}{Kulkarni, Chakraborty and Sium, et al.}




% \maketitle
\begin{abstract}
In this paper, we explore the feasibility of finding
algorithm implementations from code. Successfully matching code and algorithms can help understand unknown code, provide reference implementations, and automatically collect data for learning-based program synthesis~\cite{2019nips}. To achieve the goal, we designed a new language named {\it p-language} to specify the algorithms and a static analyzer for the {\it p-language} to automatically extract control flow, math, and natural language information from the algorithm descriptions. We embedded the output of {\it p-language} ({\it p-code}) and source code in a common vector space using self-supervised machine learning methods to match algorithm with code without any manual annotation. We developed a tool named {\it Beryllium}. It takes pseudo code as a query and returns a list of ranked code snippets that likely match the algorithm query. Our evaluation on {\it Stony Brook Algorithm Repository} and popular {\it GitHub} projects show that \toolname significantly outperformed the state-of-the-art code search tools in both C and Java. Specifically, for 98.5\%, 93.8\%, and 66.2\% queries, we found the algorithm implementations in the top 25, 10, and 1 ranked list, respectively. Given $87$ algorithm queries, we found implementations for 74 algorithms in the {\it GitHub} projects where we did not know the algorithms before.
\end{abstract}
\maketitle
\section{INTRODUCTION}
The availability of the Internet-scale open-source repositories such as {\it GitHub} provides great software reuse and sharing opportunities. The challenge for a user is how to quickly locate the right piece of code and whether to trust the software found. Consider a few concrete scenarios: (1) a researcher plans to transfer a video encoding functionality to the VLC media player~\cite{2015:Barr:ISSTA}, and wants to find a {\it donor} (some software) that implements the encoding; (2) a NASA engineer wants to run taint analysis to scan unknown Android apps; if such analysis already exists and has a good quality, {he/she} does not need to reinvent one; (3) a developer is prototyping the register allocation routine of a compiler, and believes that the graph coloring algorithm has been implemented and wishes to find a reliable and fast version to reuse or at least to reference; and (4) a self-driving car software engineer faces challenges to fix the incorrect implementations of the algorithms----the most frequently occurred bugs in their software~\cite{avbugs}, and wishes to find some reference implementations to compare against and to learn from. In all these scenarios, the user can type the names of the algorithms or the keywords of the desired functionality in the search engine, such as Google, to identify candidates; however, much more manual effort is still needed to further investigate the code before we actually can use them. Specifically, the algorithm may be implemented as part of a method without specifying the algorithm's name, and searching by the algorithm name may not work.

There have been many automatic techniques for mining software specifications in the past. Nevertheless, these specifications are either program invariants~\cite{2001:Ernst:TSE,2015:TOSEM:Yi} or the protocols of APIs~\cite{2002:Bodik:POPL,2006:Yang:ICSE}. Such specifications report code-level information and are tailored for developers rather than new software users. Code search tools use a natural language query to retrieve functions that likely match the query. For example, deep code search (DCS)~\cite{gu2018deep} trains an RNN model using code and its comments. OCoR~\cite{zhu2020ocor} proposes an attention-based neural architecture with convolutional layers to train and evaluate based on question-answer pairs from StackOverflow. Another line of work is to infer comments and summaries of the code, either via rule-based approaches~\cite{2010:Buse:ASE}, machine translations~\cite{2015:ASE:Oda}, or machine learning models~\cite{2015:ICML:Allamanis}. In both cases, code and natural language are aligned relying on supervised datasets. The techniques typically do not use code semantics, and the results can be noisy. Semantic clone detection tools can find code with similar functionalities via deep learning~\cite{deepsim,flow2vec,blend,dynamic,2021iclrgraphcodebert} or dynamic analysis~\cite{slacc}. These tools focus on comparing two implementations rather than performing a code search. They were evaluated using problem-solving competition datasets such as CodeSearchNet and Google Code Jam and have not shown to be useful for finding algorithm implementations in software repositories.

Our work aims to search open-source projects to find algorithm implementations. We use pseudo code as a search query. Pseudo code is available in textbooks and papers, typically proofread. It provides important computation steps compared to a short natural language query; thus, more details are available for accurate search. Pseudo code is system-independent and can be applied to search for different programming languages. Our search results are the code fragments that implement the algorithms located within a single method or across multiple methods and thus are more fine-grained than returning a method, as most existing search tools do.

Our approach consists of three components: {\it algorithm analysis}, {\it representation learning}, and {\it vector-based code retrieval}. The key idea is to extract control flow, maths, and natural language features from algorithms (pseudo code). We then use CodeBERT to align the natural language of an algorithm with the source code, encode math operators extracted from the algorithm and code, and use graph {representation} learning to {embed} control flows of algorithms and code. We used these three approaches to convert the algorithm and code into vectors and designed a scalable vector-based search to cluster the matched source code nodes on the {\it interprocedural control flow graphs} (ICFG). These nodes form the code fragment that matches the query. 

To enable the automatic analysis of algorithms, we developed a {\it p-language} that can specify an algorithm's control flow, natural language, and math features. We developed a learning-based method to automatically convert pseudo code to {\it p-code} (the algorithm specified in p-language). We developed a static analysis tool to construct a control flow graph for {\it p-code}, where the nodes are either math operations or natural language descriptions. We hypothesize that although the implementations may add more details when expanding from the algorithms, the main control flow of the algorithm is still in the code, and the main computation steps described in math operations and natural language in algorithms can still be mapped to the code; when the two are considered together, we can effectively distinguish one algorithm implementation from the other.

To enable the search, the source code and {\it p-code} need to be represented and compared in a common space. We used control flow graph-based embedding to map the pseudo code to the source code at the semantic level. To encode control flow, we applied the \textit{graph autoencoder (GAE)} framework and trained a model on the ICFGs of the programs via self-supervised learning. The model can generate node embeddings for the ICFGs of both {\it p-code} and source code. To encode natural language semantics in {\it p-code}, we used pre-trained CodeBERT~\cite{feng2020codebert}, which can bridge the natural language descriptions and source code. We added math encoding as another part of the node embedding. Representation learning does not require human supervision, and the trained models can generate representations for unseen pseudo code and source code.

To perform code retrieval, we consider both the semantic similarity of the code to the queried pseudo code and the connectivity of the code fragment on its ICFG. We retrieved candidate source code nodes in the first step by comparing the source code and pseudo code node-level embedding. In the second step, we grouped the candidate code nodes into code fragments by their distances on the ICFGs. The code fragments are then ranked by a score function that considers the distances of the matched nodes within the code fragment and its coverage to the pseudo code query.

We implemented our algorithm analysis, representation learning, and search algorithm in a tool called {\it Beryllium}. We collected pseudo code of 103 algorithms from the textbooks of {\it Introduction to Algorithms} \cite{2009:Cormen:Book} and {\it Algorithms Design Manual} \cite{skiena1998algorithm}. We used 10 Java and 10 C popular {\it GitHub} projects as well as the {\it Stony Brook Algorithm Repository}~\footnote{\url{http://algorist.com/algorist.html}}, which consists of $67$ Java and $27$ C real-world open-source projects that implement a known algorithm. We set up two code databases, {\it Stony Brook code database} and {\it GitHub code database} and designed three experiments that represent real-world search scenarios. We ran a total of $65$ algorithm queries for the {\it Stony Brook code database}, and $87$ queries for {\it GitHub code database}. For 64 ($98.5\%$), 61 ($93.84\%$), and 43 ($66.15\%$) queries, we found the algorithm implementations in the top 25, 10, 1 ranked list respectively, in the important projects such as {\it Guava}, {\it AlgoDS}, {\it kdtree}, and {\it Concorde} in {\it Stony Brook code database}. For 87 algorithm queries, we searched the {\it GitHub code database}, and we found $74$ algorithm implementations we did not know before. Our results show that \toolname can handle Java, C, and a mixture of C and Java languages. Our tool has outperformed the state-of-the-art tools of Deep Code Search (DCS)~\cite{gu2018deep} and OCoR~\cite{zhu2020ocor}.

In summary, the research contributions of our paper include the following: 
\begin{enumerate}
	\item the {\it p-language} and its tools to enable automatic analyses of algorithms (\S 3),
	\item the representation learning that integrates control flow, math, and natural language information of algorithms and source code (\S 4),
	\item a search algorithm that uses the embedding to locate the code fragments of algorithm implementations (\S 5),
	\item the tool, namely \toolname, that can find algorithm implementations for different programming languages (\S 6)
	\item the evaluation that demonstrates \toolname is effective for real-world code and outperformed the state-of-the-art baselines (\S 6), and
	\item the dataset~\footnote{ http://anonymous\_for\_review} of the algorithms (in both pseudo code and {\it p-code}) and the mappings between algorithms and source code fragments found in real-world Java and C projects (\S 6).
\end{enumerate}

\section{OVERVIEW}
In this section, we show an example of the input and output of \toolname and then present an overall workflow of our approach.

\subsection{An Example Input and Output}

\begin{figure}
	\begin{subfigure}[t]{0.6\columnwidth}
			\lstset{mathescape=true,breaklines=true, showlines=true,numbers = left, basicstyle = \footnotesize}
\begin{lstlisting}[style=mystyle-java]
MATRIX-MULTIPLY(A,B)
n = A.rows
let C be a new $n \times n$ matrix
for i = 1 to n
    for j = 1 to n
        $c_{ij}$ = 0
        for k = 1 to n
            $c_{ij} = c_{ij} + a_{ik} \cdot b_{kj}$
return C
\end{lstlisting}
			\caption{Input: pseudo code from {\it Introduction to Algorithms}}
			\label{fig:matrixMultPsuedoCode}
	\end{subfigure}
	\begin{subfigure}[t]{0.8\columnwidth}
		    \lstset{language=java,basicstyle=\footnotesize,numbers=left,escapeinside={<@}{@>}}
\begin{lstlisting}[style=mystyle-java]
for (int iBlock = 0; iBlock < blockRows; ++iBlock){
  final int pStart = iBlock * BLOCK_SIZE;
  final int pEnd = FastMath.min(pStart + BLOCK_SIZE,rows);
  for (int jBlock = 0; jBlock < blockColumns; ++jBlock) {
    final double[] block = blocks[iBlock * blockColumns+...;
    final int qStart = jBlock * BLOCK_SIZE;
    final int qEnd = FastMath.min(qStart+BLOCK_SIZE,...);
    int k = 0;
    for (int p = pStart; p < pEnd; ++p) {
        double sum = 0;
        int q = qStart;
        while (q < qEnd - 3) {
            sum += block[k] * v[q] +
                       block[k + 1] * v[q + 1] +
                       block[k + 2] * v[q + 2] +
                       block[k + 3] * v[q + 3];
            k += 4;
            q += 4;}
        while (q < qEnd) sum += block[k++] * v[q++];
        out[p] += sum;}}
  return out;}
\end{lstlisting}
		\caption{Output: implementation from {\it Apache Commons Math.}} 
			\label{fig:matrixMultImpl}		
	\end{subfigure}
	\caption{An Example: Matrix Multiplication}~\label{matrixmulti}
\end{figure}

\begin{figure}
\resizebox{\textwidth}{!}{
	\includegraphics{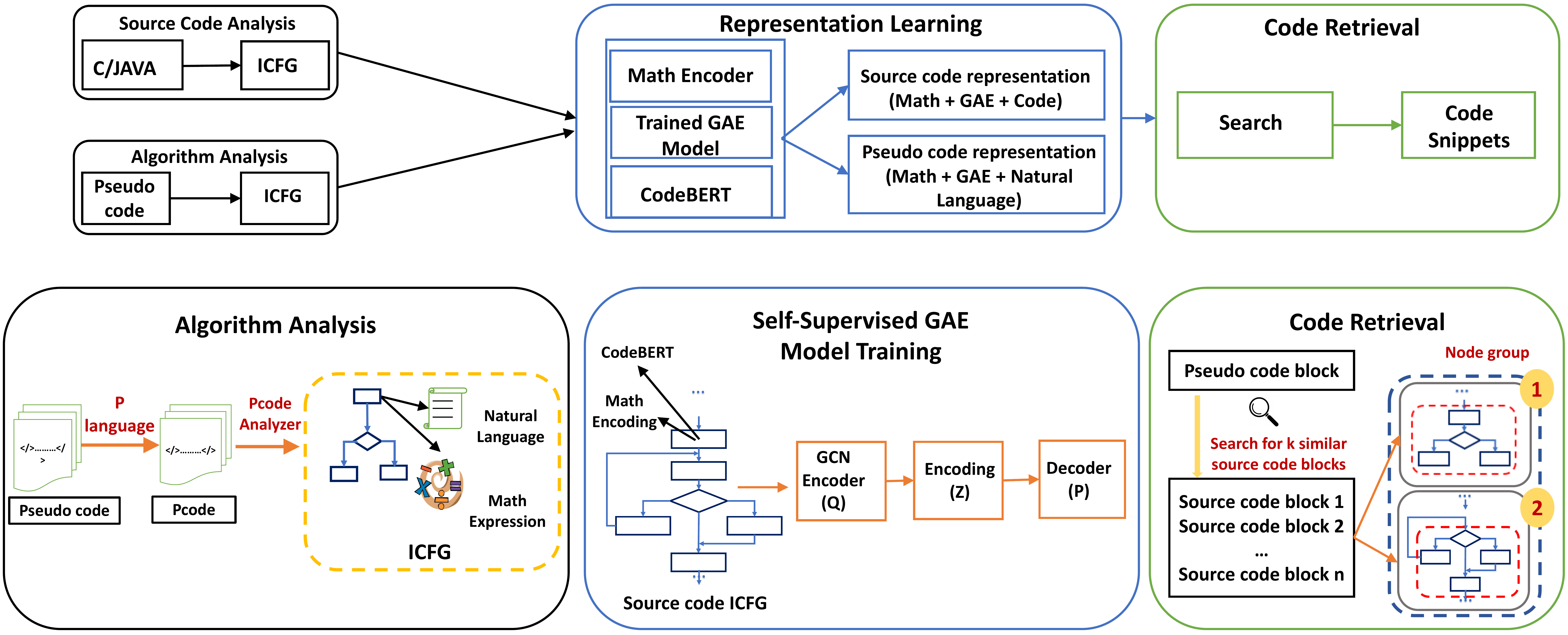}
	}
	\caption{Workflow of Beryllium}
	\label{fig:flowChart}
\end{figure} 
The matrix multiplication algorithm is commonly used in practice, e.g., in machine learning libraries. Figure~\ref{fig:matrixMultPsuedoCode} displays the pseudo code of matrix multiplication from {\it Introduction to Algorithms}~\cite{2009:Cormen:Book}. It serves as a query when we search for algorithm implementations. \toolname outputs a code snippet, instead of a line, a file, or a function, that can precisely locate the algorithm implementation from the given search codebase. \toolname captures the important commonalities between the pseudo code and implementations in the source code and is resilient to the variances caused by optimizations and data structure changes. Figure~\ref{fig:matrixMultImpl} shows the real-world implementation of matrix multiplication we found from {\it Apache Commons Math} using the query in Figure~\ref{fig:matrixMultPsuedoCode}.

The implementation of the matrix multiplication is located inside a method called {\tt operate}. Even though the implementation used the optimizations and customized data structure, we still see the control flow mappings between the algorithm and the implementation: the loops at lines~4, 5, and 7 in Figure~\ref{fig:matrixMultPsuedoCode} correspond to the loops at lines~1, 4, and 9 in Figure \ref{fig:matrixMultImpl}. The math operations at line~8 in Figure~\ref{fig:matrixMultPsuedoCode} can be found at line~13 in Figure~\ref{fig:matrixMultImpl}. The variance is that lines 12--20 in Figure~\ref{fig:matrixMultImpl} are optimized to reduce the number of loop iterations by a factor of 4. The implementation is also optimized for the cache. It stores the elements of a matrix in an array ({\tt block} at line~5 in Figure \ref{fig:matrixMultImpl}) such that when traversing the matrix during multiplication (lines~9--22), it can together bring the elements of the matrix into the cache.

\subsection{The Workflow of Our Approach}

In Figure \ref{fig:flowChart}, we show the workflow of \toolname. \toolname takes as input (1) the pseudo code of an algorithm as a search query and (2) a program (or a collection of programs) as the code search database. \toolname outputs the code snippets that implement the algorithm. A code snippet can consist of a function, selected lines in a function, or several functions located in one or multiple files.

\toolname consists of three main components. The {\it Algorithm Analysis} component (details in \S3) first translates the pseudo code to {\it p-code}, which is written in the {\it p-language} we designed, and then automatically analyzes the {\it p-code} to extract control flow, math, and natural language features from the algorithm. Similarly, the source code is analyzed to extract the control flow and math features. The {\it Representation Learning} component (presented in \S4) used a math encoder and CodeBERT to generate the node embedding for ICFGs of pseudo code and source code and then applied a trained {\it GAE} model to generate vector representations for the ICFGs. The {\it GAE} model is trained using self-supervised learning on source code based on the link prediction on the ICFGs. The {\it Code Retrieval} component (see \S5) computes the semantic similarities between the algorithm query and source code nodes and then groups the nodes based on (1) their structures in the ICFG and (2) the coverage of the algorithm query. The output is a ranked list of code fragments.
 
\section{Automatic Algorithm Analysis}
A common approach to describe an algorithm is to use pseudo code. This section explains how we automatically analyze pseudo code to extract control flow, math, and natural language features for search.

\subsection{Three Ingredients of Algorithm Descriptions}
Generally, an algorithm includes three types of information: (1) control flow of computational steps, (2) descriptions of computation in natural languages, and (3) mathematical expressions. 

According to~\cite{2009:Cormen:Book, definition}, an algorithm is defined as a sequence of unambiguous computational steps, namely {\it instructions}, that transform the input to the output. If an algorithm is well presented, we should be able to identify the orders, i.e., the control flow, of the instructions. There are three types of control flows in algorithms~\cite{2017:Xu:Tutorial,2008:Brian:notes}: {\it sequential}, {\it conditional} and {\it iterative}. In fact, {\it Introduction to Algorithms} uses the keywords {\tt while}, {\tt for}, {\tt repeat-until} and {\tt if-else} to specify the loops and conditions as its pseudo code convention. 

Unlike a program that requires rigorous syntax for compilation and execution, pseudo code is designed mainly for human comprehension. It uses a mixture of natural language descriptions and {\it mathematical expressions} --- the expressions that contain mathematical operators and symbols --- to illustrate the computation. For example, in Figure~\ref{fig:matrixMultPsuedoCode}, line~3 is a natural language description, and line~8 is a mathematical expression. 

We design the {\it p-language} that explicitly marks the control flow, natural language, and math information in pseudo code, thus making it feasible to automatically extract such information from the algorithm descriptions.

\subsection{P-Language and its Static Analyzer} \label{sec:pcode}
\label{P-Language and its Parser}
In the {\it p-language}, we introduced a set of keywords, following the pseudo code convention provided by {\it Introduction to Algorithms}, to mark the control flow of the instructions in an algorithm. The instructions are composed of {\it natural language descriptions} and {\it mathematical expressions}. 

In Figure~\ref{fig:grammar} we present the grammar of the {\it p-language}. On the left, we show that an algorithm is specified in a function consisting of a sequence of {\tt stmt}. Besides the calls and returns, the important {\tt simple\_stmt} consists of {\tt math} and {\tt natural\_language} expressions. We use a pair of `\$'s to mark the math expressions and a pair of `@'s to record the natural language descriptions. On the right of Figure~\ref{fig:grammar}, we show how the control flow of {\it branches}, {\it loops}, and {\it calls} are made explicit using a set of keywords such as `{\tt if}', `{\tt while}' and `{\tt return}'.

\begin{figure*}[!htb]
	\resizebox{\textwidth}{!}{
		\begin{minipage}{.50\linewidth}
			\begin{subfigure}[b]{\textwidth}
				\begin{description}
					\item // {\it function}
					\item [{func:}] NAME parameters suite
					\item [{parameters:}] '(' para (, para){*} ')'
					\item [{para:}] NAME $|$ expr
					\item [{suite:}] '\{' stmt+ '\}' 
				\end{description}
				\begin{description}
					\item // {\it statement}
					\item [{stmt:}] simple\_stmt $|$ compound\_stmt 
					\item [{compound\_stmt:}] if\_stmt $|$ while\_stmt $|$ repeat\_stmt $|$ for\_stmt
					\item [{simple\_stmt:}] expr $|$ call\_stmt $|$ return\_stmt
					\item [{expr:}] natural\_language $|$ math 
					\item [{math:}] '\$' {MATH EXPRESSION} '\$'
					\item [{natural\_language:}] '@' {DESCRIPTION} '@'
			 %//useful to detect 
					%recursion
				\end{description}
			\end{subfigure}
		\end{minipage}
		
		\begin{minipage}{.50\linewidth}
			\begin{subfigure}[b]{\textwidth}
				\begin{description}
					\item // {\it branch, loop and call} 
					\item [{if\_stmt:}] 'if' test suite ('elseif' 
					test suite){*} {[}'else' suite{]}
					\item [{while\_stmt:}] 'while' test suite 
					\item [{for\_stmt:}] 'for'|'for each' expr {[}'to'$|$'downto' expr{]} 
					suite 
					\item [{repeat\_stmt:}] 'repeat' suite 'until' test
					\item [{test:}] and\_test $|$ not\_test $|$ or\_test 
					\item [{and\_test:}] (test $|$ expr) 'and' (test $|$ expr) 
					\item [{not\_test:}] 'not' (test $|$ expr)
					\item [{or\_test:}]  (test $|$ expr) 'or' (test $|$ expr)
					\item [{return\_stmt:}] 'return' {[}expr $|$ call\_stmt{]}
					\item [{call\_stmt:}] NAME parameters
				
				\end{description}
			\end{subfigure}
		\end{minipage}
	}
	\caption{The grammar of the \textit{p-language}~\label{fig:grammar}}
\end{figure*}

After pseudo code is annotated using {\it p-language}, we call it {\it p-code}.
Based on the grammar, we built a static analyzer that parses the {\it p-code} and automatically generates the ICFGs for {\it p-code} based on its control flow. Similar to the ICFG of source code,  the ICFG of {\it p-code} has the nodes of the conditions, loops, and call sites. The difference is that a {\it p-code} ICFG contains the nodes of math expressions or natural language descriptions instead of source code statements. In Figure~\ref{matrixmulti}, we show two examples of {\it p-code}.

\begin{figure}
	\begin{subfigure}[t]{0.6\columnwidth}
			\lstset{mathescape = false, numbers = left, basicstyle= \footnotesize, mathescape=true}
\begin{lstlisting}[style=mystyle]
MATRIX-MULTIPLY(A,B)
$\text{\$}$n = A.rows$\text{\$}$
@let C be a new n x n matrix@ 
$\text{\bf for}$ $\text{\$}$i = 1$\text{\$}$ $\text{\bf to}$ $\text{\$}$n$\text{\$}$
    $\text{\bf for}$ $\text{\$}$j = 1$\text{\$}$ $\text{\bf to}$ $\text{\$}$n$\text{\$}$
	$\text{\$}$C[i][j] = 0$\text{\$}$
        $\text{\bf for}$ $\text{\$}$k = 1$\text{\$}$ $\text{\bf to}$ $\text{\$}$n$\text{\$}$
	      $\text{\$}$C[i][j] += a[i][k] * b[k][j]$\text{\$}$
$\text{\bf return}$ $\text{\$}$C$\text{\$}$
\end{lstlisting}
			\caption{{\it p-code} for matrix multiplication in Figure 1a} \label{fig:matrixMultPcode}
	\end{subfigure}
	\begin{subfigure}[t]{0.8\columnwidth}
		       \lstset{numbers = left, breaklines=true, basicstyle=\footnotesize, mathescape=true} %,escapeinside={<@}{@>}}
\begin{lstlisting}[style=mystyle]
MST-KRUSKAL(G,w)
$\text{\$}A=\emptyset\text{\$}$
$\text{\bf for each}$ @vertex v $\in$ G.V@ 
    MAKE-SET(v)		
@sort edges of G,E into nondecreasing order by weight w@
$\text{\bf for each}$ @edge (u,v) in G.E in nondecreasing order by weight@
$\text{\bf if}$ $\text{\$ FIND-SET(u)} \neq \text{FIND-SET(v)\$}$
       $\text{\$}A = A \cup \{(u,v)\}\text{\$}$
       UNION(u,v)
$\text{\bf return}$ $\text{\$}$A$\text{\$}$	

\end{lstlisting}
	    \caption{{\it p-code} for Kruskal, minimum spanning tree algorithm~\cite{2009:Cormen:Book}} \label{fig:kruskalPcode}
	\end{subfigure}
	\caption{{\it p-code} examples}~\label{matrixmulti}
\end{figure}
\subsection{Automatically Converting Pseudo Code to P-Code}

To convert pseudo code to {\it p-code}, we first identify the control flow structure of pseudo code, using a set of keywords defined in the {\it Introduction to Algorithms} textbook such as {\tt if}, {\tt for each}, {\tt downto}, shown in Figure~\ref{fig:grammar}. We then train a classifier to distinguish whether a statement is a math expression or a natural language description.

We use samples from source code as labeled training data to establish the training data without human annotation. Specifically, the comments are labeled as natural language descriptions, and the non-comments are labeled as math expressions. The training samples collected from the source code may have different feature distributions than the pseudo code's math expressions and natural language descriptions. For robust learning, we used the {\it label propagation algorithm}~\cite{zhou2003learning}, where the model can jointly consider the labeled data (from source code) and the distribution of unlabeled data (from pseudo code). Using this approach, we construct a network using all the data points (including both labeled and unlabeled); the similarity of each pair of data points is then marked on the edges of the network. The labels get propagated from the labeled data to unlabeled data via a random walk of the network until all the data are labeled.

\section{Representation of Code and Algorithms}
\label{Representation of Code and Algorithms}
In this section, we describe how to learn the representation of the {\it p-code} ICFGs and source code ICFGs to make them comparable for search. At a high level, we first used a collection of source code ICFGs to train a graph neural network {\it (GNN)}. The model training is self-supervised, and no human annotation is needed. Once the model is trained,  we then use it to generate vector representations for any input {\it p-code} query and new source code search database.

\subsection{Comparing P-Code and Source Code}
Using {\it p-code}, we can automatically extract three important features for algorithms, namely {\it control flow}, {\it math expressions}, and {\it natural language descriptions}. Control flow is an important feature because we believe that the logic of branches, loops, and the sequential order of major computation steps in the pseudo code will likely be preserved in its implementation. In fact, algorithmic complexity is computed using control flow information like loops and recursive calls in the pseudo code and is widely used to compare and estimate running software performance. On the other hand, the control flow of pseudo code and the corresponding source code implementation can have differences at a low level. For example, a math operator $\Sigma$ in the pseudo code may be translated into a loop to compute a summation of a set of integers. Thus, we cannot use the exact match of the ICFGs between the {\it p-code} and source code. We apply {\it GNN} to the ICFGs and compare the ICFG nodes of {\it p-code} and source code using generated vector representations.

One key characteristic of {\it p-code} is that it contains a mixture of natural language description and math expressions. To effectively align the two pieces of information with source code for comparison, we used {\it math encoding} for representing math expressions for both {\it p-code} and source code and used CodeBERT to encode the natural languages of {\it p-code} and code from the source code. CodeBERT is a bi-model BERT pre-trained in both natural language and programming languages \cite{feng2020codebert}. For {\it p-code}, the statements between `@'s in {\it p-code} (natural language description nodes) are encoded using the natural language component of CodeBERT. For source code, we use the code component of CodeBERT to get the encoding for all nodes. Since CodeBERT was pre-trained using aligned natural languages (comments) and programming languages (source code), the CodeBERT encoding for the natural language of {\it p-code} and source code are comparable.

To generate math encoding for {\it p-code} and source code, we establish a mapping from the math operators in C/Java/pseudo code to 7 types, including {\it AddSub} ('+', `-'), {\it MultDiv} (`$*$', `/'), {\it Deference} (`[', `]'), {\it Modular} (`\%'), {\it Bit Operator} (`<<', `>>'), {\it Logical Operator} (`\&\&', `!', `||') and {\it Relational Operator} (`>=', `<=', `<', `>', `!=', `=='). Note that we group the complement types into one category, e.g.,`+' and `-' are in the same group, as $1+2$ and $1-(-2)$ are semantically identical. We then count the frequency of such operators in the math expressions of {\it p-code} (statements between `\$'s) and in the source code.

CodeBERT and math encoding are computed for each node in the ICFGs of {\it p-code} and source code. The ICFGs with such node embeddings are then sent to the trained {\it GNN} models to generate the final node representations for the ICFGs of {\it p-code} and source code. In the following section, we provide details on how to train such a model.

\subsection{Training a Model for ICFGs via Self-Supervised Learning}\label{sec:gnnembedding} 

We applied the {\it graph autoencoder} ({\it GAE}) framework to train a model to encode the control flow structure of both {\it p-code} and source code. The model training is self-supervised and requires no human annotations. {\it GAE} has an encoder and decoder type of architecture, and specifically, it consists of an encoder $Q(\boldsymbol Z|\boldsymbol A, \boldsymbol X)$, and a decoder $P(\boldsymbol A|\boldsymbol Z)$, where $Z$ is the encoding matrix, $\boldsymbol A$ is the adjacency matrix of the graph ($A_{ij}$=1 if there is an edge between nodes $i$ and $j$; otherwise, $A_{ij}$=0), and $\boldsymbol X$ is the feature matrix ($X_{i}$ is the embedding for node $i$). In the {\it GAE} framework, we used {\it Graph Convolutional Network (GCN)} as an encoder. The $\ell^{th}$ {\it GCN} layer is defined as:
\begin{align}
    \boldsymbol Z^{(\ell+1)} = ReLU\left({\boldsymbol D}^{-\frac{1}{2}} \boldsymbol A {\boldsymbol D}^{-\frac{1}{2}} \boldsymbol X^{(\ell)} \boldsymbol W^{(\ell)}\right),
\end{align}
where, $\boldsymbol A$ is the adjacency matrix, $\boldsymbol D$ is the degree matrix (where $D_{ii}$ lists the number of the neighbours of node $i$ and $D_{ij}$=0 when $i\neq j$), $\boldsymbol X^{(\ell)}$ is the input feature matrix at layer $\ell$, $\boldsymbol W^{(\ell)}$ is the trainable parameter matrix, and $ReLU(\cdot)$ represents the ReLU activation function. Given the final layer of node encoding $\boldsymbol Z$, the goal of the decoder is to reconstruct the original graph by predicting the relationship between the nodes. The decoder uses a dot product to predict if an edge exists between two nodes~\cite{kipf2016variational}:
\begin{align}
    \boldsymbol P(\boldsymbol{\hat{A}}_{i,j}|z_i, z_j)= \sigma(z^T_i  z_j),
\end{align}
where, $\boldsymbol{\hat{A}}_{i,j}$ is the predicted edge between node $i$ and node $j$, and  $\sigma(\cdot)$ is logistic sigmoid function.

In self-supervised learning in {\it GAE}, we consider the actual edges between the nodes as the {\it positive samples}. We sample an equal number of non-existing edges between randomly selected nodes as the {\it negative samples}. We use {\it GAE} to predict those edges during training and then compare them with the actual graph. The training minimizes the cross-entropy loss between the reconstructed adjacency matrix and the true adjacency matrix:
\begin{align}
    \boldsymbol L =\boldsymbol \sum_{i\in V, j\in V} (-A_{i,j}\log\hat{A}_{i,j} - (1 - A_{i,j})log(1 - \hat{A}_{i,j})).
\end{align}

The {\it GAE} is trained using the code database. We then use the trained model to generate the ICFG node embedding for any given source code and {\it p-code}:
\begin{align}
    \boldsymbol Z =\boldsymbol p_Q(X, A),
\end{align}
where, Z is node embedding, $p_Q$ is the {\it GCN} encoder, $X$ is the initial feature vector, and $A$ is the adjacency matrix of the graph. In such a way, the model can handle any queried algorithms for a code search database. 
Finally, the {\it GCN} generated embedding is concatenated with the initial features for each node as the final representation of the node. 

\section{Code Retrieval}
In this section, we present our search algorithm. First, we describe our approach of mapping the {\it p-code}  nodes to the source code nodes in the code database in Section \ref{matching_nodes}. Then, we explain our {\it node grouping} method, which is the process of grouping matched source code nodes to find the code fragments for the pseudo code query in Section~\ref{node_grouping}.

\subsection{Matching Nodes}
\label{matching_nodes}
\begin{figure}
	\includegraphics[width=0.8\textwidth]{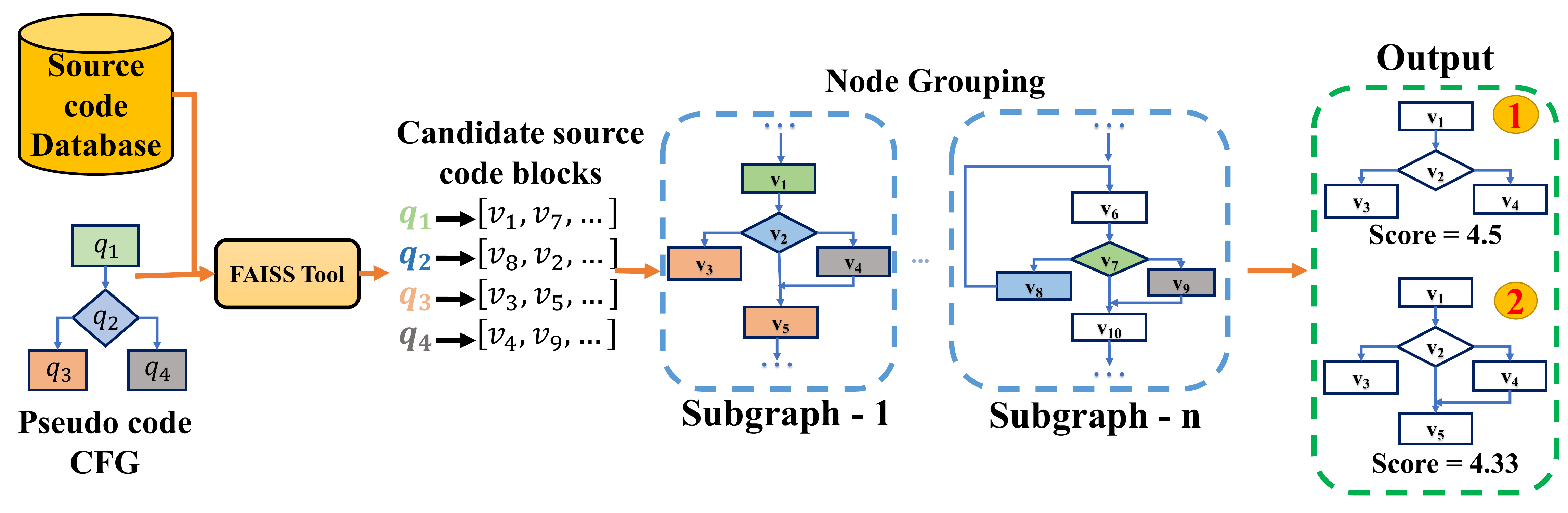}
\vspace{-0.1in}
	\caption{Node Matching and Grouping}
	\label{fig:search}
\end{figure}

As shown in Figure~\ref{fig:search}, we take one pseudo code query and the source code database we aim to perform the search.
First, the pseudo code is converted to {\it p-code}, and the ICFG is generated based on the {\it p-code} (Section \ref{P-Language and its Parser}). After the source code and {\it p-code} have been represented in vectors (Section \ref{Representation of Code and Algorithms}), we compute the cosine similarity ($S_C$) between each node $i$ in the {\it p-code} ICFG ($G^{P}$), whose node embedding is $e_{i}^{P}$, and the encoding $e_{j}^{S}$ of each node $j$ from the source code ICFGs ($G^{S}$) as follows:
\begin{equation}
    S_C(e_{i}^{P}, e_{j}^{S}) = \frac{e_{i}^{P}\ \cdot\ e_{j}^{S}}{||e_{i}^{P}||\ ||e_{j}^{S}||},
\end{equation}
where, the numerator is the dot product between the encodings and $||\cdot||$ represents the euclidean distance of the vector to the origin. Cosine similarity measures the angle of the two vectors and is widely used to measure semantic similarity in vector space. 

We applied the FAISS tool~\footnote{\url{https://github.com/facebookresearch/faiss}} to index the source code database for efficient search and comparison. In the output of the FAISS tool, the nodes in the source code ICFGs are sorted based on the cosine similarity, and the top $k$ are chosen as the candidate source code nodes. For example, in Figure~\ref{fig:search}, node $q_1$ in pseudo code is matched to $v_1$, $v_7$ ... in the source code database based on their cosine similarity. 

\subsection{Node Grouping}
\label{node_grouping}

The search algorithm aims to find code fragments from the source code database that implement the algorithms. In addition to node-level similarity, the matched source code nodes should also be close to each other on the source code ICFG to form a high-quality code fragment. Based on this idea, we develop a node grouping method using the {\it agglomerative hierarchical clustering}~\cite{han2011data} algorithm. It is a greedy technique that gradually groups the nodes with the shortest distance until all the matched nodes are grouped into the code fragment. Specifically, the input to node grouping is the matched source code nodes from the previous step. Initially, each node is considered as a separate group, and then groups with minimal distances are merged, where the distance is calculated as follows:
\begin{equation}
    distance = \max\{sp(x, y): x \in {C_1}, y \in {C_2}\}
\end{equation}
$sp(x,y)$ is the shortest path on the ICFG between nodes $x$ and $y$ which belong to groups ${C_1}$ and ${C_2}$, respectively. Note that if there is no path from node $x$ to node $y$ on the graph, the distance is $\infty$. If the distance of two groups is $\infty$, the two groups will not be merged.

The groups obtained after each merging step are considered code fragment candidates. We design a scoring function that considers two factors: (1) the coverage of the group in the {\it p-code} query, and (2)
the distances of the code blocks from the same group in the source code ICFG. Intuitively, the group with more covered nodes in the {\it p-code} query and with code blocks closer to each other in the source code ICFG should be ranked higher and thus should have a higher score. The confidence score, $\gamma$, for each candidate group ($\mathbb{C}$), is computed as follows:
\begin{equation}
    \gamma = QC + \frac{1}{distance'}
\end{equation}
where $QC$ refers to the number of covered nodes in the {\it p-code} query, and $distance' = \max \{ sp(x,y): x, y \in \mathbb{C} \}$ referring to the maximum shortest path between any two nodes within a group.

\section{Evaluation}
In our evaluation, we investigated the following research questions:

\begin{itemize}

\item \textbf{RQ1:} Can \toolname effectively retrieve the algorithm implementations from code?

\item \textbf{RQ2:} How is our approach compared to the baselines?

\item \textbf{RQ3:} Does our approach work for different programming languages?

\item \textbf{RQ4:} Which features are the most useful for search? 
\end{itemize}

\subsection{Experimental Setups and Implementation}

\subsubsection{Subjects} We collected the pseudo code for a total of $103$ algorithms from {\it Introduction to Algorithms} \cite{2009:Cormen:Book} and {\it Algorithms Design Manual} \cite{skiena1998algorithm}. We used the {\it Stony Brook Algorithm Repository}~\footnote{\url{http://algorist.com/algorist.html}} to construct a code database which we call the {\it Stony Brook code database}. We chose this repository because it provides the ground truth and documents the algorithm(s) implemented in the project. However, it did not provide where the algorithms are located in the code. This repository contains $67$ C and $27$ Java real-world open-source projects (a total of 5.2 million lines of code), among which $51$ C and $22$ Java projects can be processed by the ICFG construction tool Atlas~\footnote{\url{https://www.ensoftcorp.com/atlas/sdk/}}, which we used. These projects implemented a total of $29$ algorithms out of the $103$ algorithms we collected, where $29$ algorithms have Java implementations, and $18$ algorithms have C implementations.

\begin{figure}
 	\includegraphics[width=0.99\textwidth]{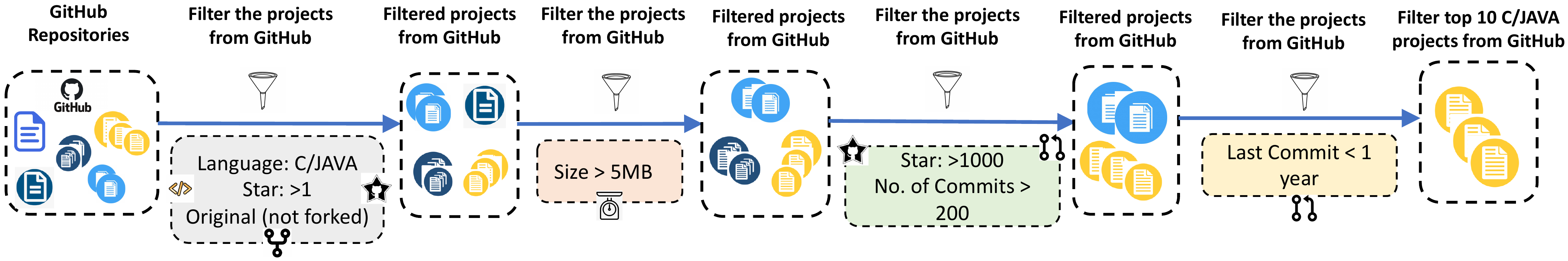}
 	\caption{Flowchart for GitHub project selection }
  \label{fig:Flowchart}
 \end{figure}

\begin{figure}
	\includegraphics[width=0.99\textwidth]{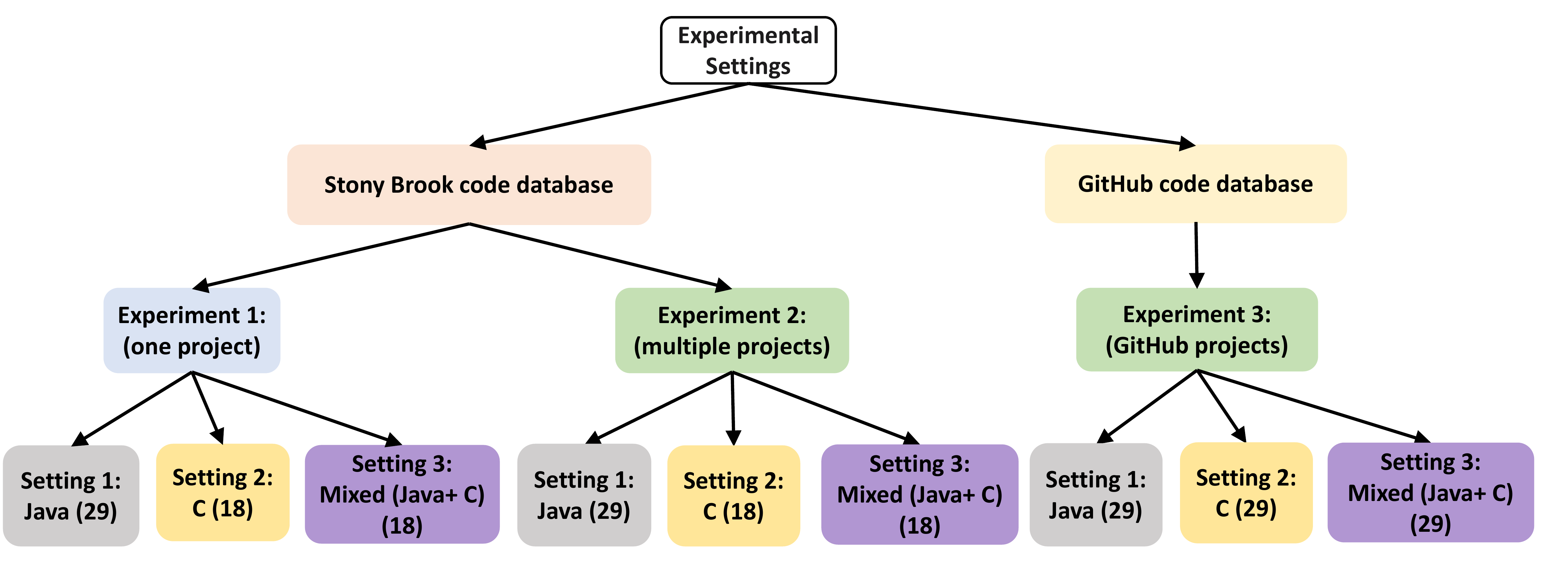}
	\caption{Organization of Experiments. Numbers in parentheses indicate the number of queried algorithms}
 \label{fig:Flowchart_exp}
\end{figure}

In addition to the ground truth projects, we included frequently used important open-source projects and tested if our tool could be useful in practice. To do so, we constructed another code search database which we call the {\it GitHub code database} by collecting $10$ C and $10$ Java popular projects from {\it GitHub} that (1) are original projects (not forked), (2) have a size greater than 5MB, (3) have recent commits, (4) have stars greater than 1000 on {\it GitHub}, and (5) can be compiled with Atlas to generate the ICFGs. The flowchart of the {\it GitHub} project selection is shown in Figure \ref{fig:Flowchart}. We used the same $29$ algorithms for the experiments on the {\it GitHub code database} as the ones used for {\it Stony Brook code database}. This makes confirmation of the ranked algorithms implementation easier and more scalable. The organization of the experiments is shown in Figure \ref{fig:Flowchart_exp}.

\subsubsection{Implementation} We implemented  \toolname that includes the following components: 

{\noindent \bf Converting pseudo code to p-code:} To establish the train set, we randomly collected $20$ lines of comments and $712$ lines of source code from Java implementations as the labeled data and collected $1023$ lines of pseudo code from the textbooks of {\it Introduction to Algorithms} \cite{2009:Cormen:Book} and {\it Algorithms Design Manual} \cite{skiena1998algorithm} as the unlabeled data. We used the label propagation algorithm implementation by Scikit-learn library~\footnote{\url{https://scikit-learn.org/stable/index.html}} to train the classifier and predict on the unlabeled pseudo code.   

{\noindent \bf Generating ICFGs for source code and p-code:} The {\it p-code} analyzer is implemented in C language based on the {\it p-language} defined in Section~\ref{sec:pcode} and is used to convert the {\it p-code} to the ICFGs. We used Atlas to generate source code ICFGs for both C and Java language source code.

{\noindent \bf GAE training}: 
We trained three {\it GAE} models, one using Java code, one using C code, and one using both C and Java code. All the training data are from the Stony Brook code database. Note that we used self-supervised learning of link prediction on the ICFGs, so no human labels are required. We implemented {\it GAE} using the {\it PyTorch Geometric library}. We used two layers of {\it GCN} as encoder, and each layer's output dimension is $512$. For all the experimental settings, the data set is split into the ratio of 8:2 as training and validation sets, respectively. The hyperparameters are tuned on the validation set, and the training is terminated when the performance on the validation set does not improve. We used Adam optimizer without dropout, the learning rate is $\eta = 0.01$, and the batch size is $2048$. We measured the performance of the trained model in terms of the area under the ROC curve (AUC).

{\noindent \bf Search Component}:  We implemented the search component in Python and used the FAISS tool as a subcomponent. We replaced the Euclidean distance similarity metrics with the cosine similarity defined in Section \ref{matching_nodes}. We obtained the top $100$ matching blocks from the code search database for each queried {\it p-code} node for a given query. 

\subsubsection{Experiment Design}
\label{experiment design}
We conducted experiments on two code databases, the {\it Stony Brook} and {\it GitHub code databases} for three language settings: Java, C, and a mixture of Java and C. The organization of experiments is illustrated in Fig.~\ref{fig:Flowchart_exp}. We designed two experiments for the {\it Stony Brook code database}, which involve three different language settings. Experiment 1 aims to determine how effectively we can find an algorithm in one project (refer to ``\textit{one project}" in the results tables). This case represents the use scenario where Alice wants to investigate a particular library or legacy project. {\it Stony Brook code database} provides the ground truth of which algorithm (s) has been implemented in a project. This helps us to better evaluate the correctness of our search results. Experiment 2 aims to determine how effectively we can find an algorithm among a set of projects (refer to ``\textit{multiple projects}" in the results tables) and how often we mistakenly identify non-relevant code as the algorithm implementation. This experiment simulates the use case where Alice wants to search for an algorithm in the company codebase where many projects exist. For the {\it GitHub code database}, we conducted one experiment (which we refer to as Experiment 3) for all three language settings. This experiment is similar to Experiment 2 of the {\it Stony Brook code database}, where we use real-world {\it GitHub} projects (refer to ``\textit{GitHub projects}" in the results tables). This experiment represents the use scenario where Alice wants to search for an algorithm among a list of popular {\it GitHub} projects.

\subsubsection{Baselines}

In the first step, we tried the na\"ive searching, i.e., using the algorithm name as a query, as a baseline. Specifically, we searched with the algorithm names as queries to find the method that implements the algorithms in all the code files from the {\it Stony Brook code database} (Experiment 2) and {\it GitHub code database} (Experiment 3). If a match is found, we retrieved the method in the code file as the method that implements the algorithm. The code files contain the code including the comments provided by the developer. For the experiments on {\it Stony Brook code database}, for the setting of Java language, the search only successfully retrieved the method for $14$ out of $29$ queries. For the setting of C language, the search only successfully retrieved the method for $3$ out of $18$ queries, and for the setting of mixed language, the search only successfully retrieved the method for $11$ out of $18$ queries. The results of Experiment 1 on {\it Stony Brook code database} are same as the results of Experiment 2 since {\it Stony Brook code database} provides ground truth and if the search retrieves the method then it is in the ground truth projects. For the experiments on {\it GitHub code database}, the search only successfully retrieved the method for $9$, $7$, and $10$ out of $29$ queries for the setting of Java, C, and mixed languages, respectively.

The na\"ive search does not work well because it solely relies on text matching. The search fails to retrieve the method if the method names, variable names or the comments in the code file do not align with the algorithm's names. An algorithm can be complex and it can be implemented using multiple methods. Therefore, the algorithm may not have a one-to-one mapping to method. Due to this, even though the na\"ive search retrieves the method, the method may not have the complete implementation of the algorithm. Due to these restrictions, na\"ive searching did not retrieve the correct methods for most algorithm queries.

In the second step, we collected the most recent work on code search that (1) can address our problem of retrieving implementation code for a queried algorithm and 
(2) has provided a {\it GitHub} repository and instructions to reproduce their approaches. 
After we tried several recent works, we found that deep code search (DCS)~\cite{gu2018deep} and OCoR~\cite{zhu2020ocor} are the SOTA methods that can satisfy the two criteria. Both of the approaches used deep learning for code search. We re-train the Java, C, and mixed language models for both baselines, using their instructions and code. To implement DCS on our data, we first process our data to extract method names, tokens from the method body that does not include Java keywords, API sequences (code provided by the authors of DCS), and comments of the code snippet for each code file. Furthermore, we split the extracted method names and tokens based on the camel case. We used the PyTorch code from their {\it GitHub} repository~\footnote{\url{https://github.com/guxd/deep-code-search}} to re-train the DCS model. To implement OCoR on our data, we process our data to obtain comments for each method and generate comment-code pairs to be used as the question-code pairs for training and validation. We used the code from their {\it GitHub} repository~\footnote{\url{https://github.com/pkuzqh/OCoR}} to train the model on our dataset. %Finally, we use the trained model to
To compare with \toolname, we used pseudo code as queries for the baselines. We have also tried algorithm names as queries; however, the baselines performed poorly.

\subsubsection{The Search Results Confirmation and Evaluation Metrics}
The output of \toolname is a ranked list of the code snippets, and the output of the baselines is a ranked list of the methods. The authors first inspected some sample outputs and established that the code snippet is confirmed to implement an algorithm if 1) the control flow of the code (the conditions and loops) aligns with the pseudo code, and 2) variables and math operators in the code match with the pseudo code. Since the implementation may not completely follow the pseudo code, the annotators also try to understand the code to determine if the algorithm is implemented. Two authors independently annotated the top 100 outputs for each query and discussed any disagreements. The code fragments are marked as unknown if the authors cannot resolve the disagreement. For baselines, if a retrieved method contains the implementation, the method is marked as a hit. Note that this is a relaxed standard compared with the standard for \toolname since the method can also contain much code irrelevant to the queried algorithm. 

With the annotation, we use {\it F-rank} to measure the tool's success. F-rank is the rank of the first hit result in the output list. If the top 100 results do not contain a hit, we denote it as ``>100''. In Experiment 1, an algorithm implementation can be returned from multiple projects. We conducted experiments on these projects one at a time and then reported the query's mean F-rank. 
We also reported MRR, the mean of inverse F-rank on all queries. For F-rank, the lower, the better, and for MRR, the higher, the better. F-rank is reported in the tabulars in all the tables, and MRR is reported in the last row of the tables.

\subsubsection{Ablation studies}
\label{ablation studies}

We conducted ablation studies to investigate the usefulness of the three features: control flow graph, math, and natural language/source code. Specifically,
we compared search results under three configurations: the CB (CodeBERT) configuration only considered source code and natural language features for search. CB (CodeBert)+G (GAE) used source code/natural language with control flow graph encoding. Finally, the results from {\it Beryllium} used all three features.

\subsubsection{Systems that conducted experiments} To convert pseudo code to {\it p-code} and to generate the ICFGs for source code and {\it p-code}, we used a MacBook Pro with 4GB AMD GPU, 16 GB RAM, and 6 CPU Cores. We used Google Colab with V100 GPU, 53 GB RAM, and 8 CPU Cores for GAE training and code retrieval.

\subsection{Results}~\label{sec:1}
\subsubsection{Stony Brook code database}
Tables~\ref{java},~\ref{c} and~\ref{mixed} display our results for the settings of Java, C, and mixed languages, respectively, on the Stony Brook code database. Each table lists the results for {\it one project} and {\it multiple projects} (described in Section \ref{experiment design}). Under {\it Query} in the tables, we listed the algorithms with known implementations in the {\it Stony Brook Algorithm Repository}. Specifically, $29$ and $18$ algorithms are implemented in the Stony Brook code database for Java and C language, respectively. There are $18$ algorithms that have both C and Java implementations (Table~\ref{mixed}). Columns DCS and OCoR reported our baseline results, and Columns {\it CB+G} and {\it CB} provided results for our ablation studies (details in Section \ref{ablation studies}). 

\begin{table}[ht]
\centering

\caption{The F-rank results for experiments on {\it Stony Brook code database} for the setting of Java language. For {\it one project}, the results show the mean over all the ground truth projects implementing the algorithm. The best performing MRR result is shown in bold.}
\label{java}
\resizebox{.96\textwidth}{!}{
\begin{tabular}{c||c||ccccc||ccccc}
\hline
             & \textbf{}                         & \multicolumn{5}{c||}{\textbf{one project}}                                                                                                                           & \multicolumn{5}{c}{\textbf{multiple projects}}                                                                                                                     \\ \hline
\textbf{No.} & \textbf{Query}                    & \multicolumn{1}{c|}{\textbf{Beryllium}} & \multicolumn{1}{c|}{\textbf{DCS}} & \multicolumn{1}{c|}{\textbf{OCoR}} & \multicolumn{1}{c|}{\textbf{CB+G}} & \textbf{CB} & \multicolumn{1}{c|}{\textbf{Beryllium}} & \multicolumn{1}{c|}{\textbf{DCS}} & \multicolumn{1}{c|}{\textbf{OCoR}} & \multicolumn{1}{c|}{\textbf{CB+G}} & \textbf{CB} \\ \hline
1            & any-segments-intersect            & \multicolumn{1}{c|}{1}                  & \multicolumn{1}{c|}{3.5}          & \multicolumn{1}{c|}{4.5}           & \multicolumn{1}{c|}{1.5}           & 1.5         & \multicolumn{1}{c|}{1}                  & \multicolumn{1}{c|}{4}            & \multicolumn{1}{c|}{4}             & \multicolumn{1}{c|}{1}             & 2           \\ \hline
2            & approx-vertex-cover               & \multicolumn{1}{c|}{5}                  & \multicolumn{1}{c|}{5}            & \multicolumn{1}{c|}{7}             & \multicolumn{1}{c|}{6}             & 9           & \multicolumn{1}{c|}{11}                 & \multicolumn{1}{c|}{7}            & \multicolumn{1}{c|}{14}            & \multicolumn{1}{c|}{12}            & 13          \\ \hline
3            & breadth-first-search              & \multicolumn{1}{c|}{2.25}               & \multicolumn{1}{c|}{6}            & \multicolumn{1}{c|}{3}             & \multicolumn{1}{c|}{5}             & 11.5        & \multicolumn{1}{c|}{9}                  & \multicolumn{1}{c|}{13}           & \multicolumn{1}{c|}{10}            & \multicolumn{1}{c|}{9}             & 21          \\ \hline
4            & compute-transition-function       & \multicolumn{1}{c|}{1}                  & \multicolumn{1}{c|}{9}            & \multicolumn{1}{c|}{5}             & \multicolumn{1}{c|}{1}             & 2           & \multicolumn{1}{c|}{1}                  & \multicolumn{1}{c|}{11}           & \multicolumn{1}{c|}{9}             & \multicolumn{1}{c|}{1}             & 1           \\ \hline
5            & extend-shortest-paths             & \multicolumn{1}{c|}{1.33}               & \multicolumn{1}{c|}{3.33}         & \multicolumn{1}{c|}{2.33}          & \multicolumn{1}{c|}{1.33}          & 1.33        & \multicolumn{1}{c|}{1}                  & \multicolumn{1}{c|}{3}            & \multicolumn{1}{c|}{2}             & \multicolumn{1}{c|}{1}             & 1           \\ \hline
6            & finite-automaton-matcher          & \multicolumn{1}{c|}{2}                  & \multicolumn{1}{c|}{12}           & \multicolumn{1}{c|}{6}             & \multicolumn{1}{c|}{2}             & 2           & \multicolumn{1}{c|}{5}                  & \multicolumn{1}{c|}{15}           & \multicolumn{1}{c|}{9}             & \multicolumn{1}{c|}{6}             & 12          \\ \hline
7            & floyd-warshall                    & \multicolumn{1}{c|}{1}                  & \multicolumn{1}{c|}{12.33}        & \multicolumn{1}{c|}{4}             & \multicolumn{1}{c|}{1.33}          & 3           & \multicolumn{1}{c|}{1}                  & \multicolumn{1}{c|}{21}           & \multicolumn{1}{c|}{14}            & \multicolumn{1}{c|}{2}             & 4           \\ \hline
8            & graham-scan                       & \multicolumn{1}{c|}{1.5}                & \multicolumn{1}{c|}{1.5}          & \multicolumn{1}{c|}{2.5}           & \multicolumn{1}{c|}{2}             & 3.5         & \multicolumn{1}{c|}{1}                  & \multicolumn{1}{c|}{2}            & \multicolumn{1}{c|}{3}             & \multicolumn{1}{c|}{3}             & 5           \\ \hline
9            & hopcroft-karp                     & \multicolumn{1}{c|}{1.5}                & \multicolumn{1}{c|}{4}            & \multicolumn{1}{c|}{8.5}           & \multicolumn{1}{c|}{2.5}           & 4           & \multicolumn{1}{c|}{7}                  & \multicolumn{1}{c|}{7}            & \multicolumn{1}{c|}{10}            & \multicolumn{1}{c|}{9}             & 15          \\ \hline
10           & insertionsort                     & \multicolumn{1}{c|}{1}                  & \multicolumn{1}{c|}{3}            & \multicolumn{1}{c|}{5}             & \multicolumn{1}{c|}{1}             & 1           & \multicolumn{1}{c|}{8}                  & \multicolumn{1}{c|}{4}            & \multicolumn{1}{c|}{7}             & \multicolumn{1}{c|}{8}             & 16          \\ \hline
11           & johnson                           & \multicolumn{1}{c|}{1.33}               & \multicolumn{1}{c|}{10.33}        & \multicolumn{1}{c|}{1.33}          & \multicolumn{1}{c|}{1.33}          & 1.33        & \multicolumn{1}{c|}{1}                  & \multicolumn{1}{c|}{12}           & \multicolumn{1}{c|}{2}             & \multicolumn{1}{c|}{3}             & 5           \\ \hline
12           & kruskal                           & \multicolumn{1}{c|}{1}                  & \multicolumn{1}{c|}{17.75}        & \multicolumn{1}{c|}{14.75}         & \multicolumn{1}{c|}{1.25}          & 1.5         & \multicolumn{1}{c|}{1}                  & \multicolumn{1}{c|}{23}           & \multicolumn{1}{c|}{18}            & \multicolumn{1}{c|}{2}             & 9           \\ \hline
13           & lcs-length                        & \multicolumn{1}{c|}{1}                  & \multicolumn{1}{c|}{3}            & \multicolumn{1}{c|}{1}             & \multicolumn{1}{c|}{2}             & 10          & \multicolumn{1}{c|}{1}                  & \multicolumn{1}{c|}{5}            & \multicolumn{1}{c|}{1}             & \multicolumn{1}{c|}{1}             & 2           \\ \hline
14           & lu-decomposition                  & \multicolumn{1}{c|}{1}                  & \multicolumn{1}{c|}{1}            & \multicolumn{1}{c|}{2}             & \multicolumn{1}{c|}{1}             & 2           & \multicolumn{1}{c|}{16}                 & \multicolumn{1}{c|}{2}            & \multicolumn{1}{c|}{6}             & \multicolumn{1}{c|}{16}            & 18          \\ \hline
15           & maybe-mst                         & \multicolumn{1}{c|}{1.2}                & \multicolumn{1}{c|}{11.4}         & \multicolumn{1}{c|}{10.2}          & \multicolumn{1}{c|}{1.2}           & 1.4         & \multicolumn{1}{c|}{1}                  & \multicolumn{1}{c|}{17}           & \multicolumn{1}{c|}{15}            & \multicolumn{1}{c|}{3}             & 17          \\ \hline
16           & modular-linear-equation-solver    & \multicolumn{1}{c|}{1}                  & \multicolumn{1}{c|}{3}            & \multicolumn{1}{c|}{3}             & \multicolumn{1}{c|}{1}             & 1           & \multicolumn{1}{c|}{1}                  & \multicolumn{1}{c|}{5}            & \multicolumn{1}{c|}{6}             & \multicolumn{1}{c|}{2}             & 13          \\ \hline
17           & mst                               & \multicolumn{1}{c|}{1}                  & \multicolumn{1}{c|}{1.2}          & \multicolumn{1}{c|}{1.2}           & \multicolumn{1}{c|}{1.2}           & 1.4         & \multicolumn{1}{c|}{1}                  & \multicolumn{1}{c|}{1}            & \multicolumn{1}{c|}{1}             & \multicolumn{1}{c|}{1}             & 1           \\ \hline
18           & mst-reduce                        & \multicolumn{1}{c|}{1.2}                & \multicolumn{1}{c|}{6}            & \multicolumn{1}{c|}{2.2}           & \multicolumn{1}{c|}{1.2}           & 1.4         & \multicolumn{1}{c|}{1}                  & \multicolumn{1}{c|}{14}           & \multicolumn{1}{c|}{4}             & \multicolumn{1}{c|}{1}             & 1           \\ \hline
19           & naïve-string-matcher              & \multicolumn{1}{c|}{1}                  & \multicolumn{1}{c|}{4}            & \multicolumn{1}{c|}{3}             & \multicolumn{1}{c|}{1}             & 1           & \multicolumn{1}{c|}{12}                 & \multicolumn{1}{c|}{6}            & \multicolumn{1}{c|}{4}             & \multicolumn{1}{c|}{12}            & 12          \\ \hline
20           & optimal-bst                       & \multicolumn{1}{c|}{1.25}               & \multicolumn{1}{c|}{9.25}         & \multicolumn{1}{c|}{12}            & \multicolumn{1}{c|}{4}             & 18.5        & \multicolumn{1}{c|}{5}                  & \multicolumn{1}{c|}{14}           & \multicolumn{1}{c|}{17}            & \multicolumn{1}{c|}{5}             & 28          \\ \hline
21           & pivot                             & \multicolumn{1}{c|}{1}                  & \multicolumn{1}{c|}{2}            & \multicolumn{1}{c|}{3}             & \multicolumn{1}{c|}{1}             & 1           & \multicolumn{1}{c|}{1}                  & \multicolumn{1}{c|}{3}            & \multicolumn{1}{c|}{7}             & \multicolumn{1}{c|}{2}             & 10          \\ \hline
22           & prim                              & \multicolumn{1}{c|}{1}                  & \multicolumn{1}{c|}{2.25}         & \multicolumn{1}{c|}{3.25}          & \multicolumn{1}{c|}{2.25}          & 3.5         & \multicolumn{1}{c|}{1}                  & \multicolumn{1}{c|}{3}            & \multicolumn{1}{c|}{3}             & \multicolumn{1}{c|}{1}             & 5           \\ \hline
23           & print\_all\_pair\_shortest\_paths & \multicolumn{1}{c|}{1.5}                & \multicolumn{1}{c|}{4}            & \multicolumn{1}{c|}{9.5}           & \multicolumn{1}{c|}{3.5}           & 6           & \multicolumn{1}{c|}{2}                  & \multicolumn{1}{c|}{7}            & \multicolumn{1}{c|}{10}            & \multicolumn{1}{c|}{3}             & 6           \\ \hline
24           & quicksort                         & \multicolumn{1}{c|}{1}                  & \multicolumn{1}{c|}{5}            & \multicolumn{1}{c|}{4}             & \multicolumn{1}{c|}{4}             & 8           & \multicolumn{1}{c|}{49}                 & \multicolumn{1}{c|}{51}           & \multicolumn{1}{c|}{55}            & \multicolumn{1}{c|}{51}            & 63          \\ \hline
25           & rabin-karp-matcher                & \multicolumn{1}{c|}{1}                  & \multicolumn{1}{c|}{5}            & \multicolumn{1}{c|}{3}             & \multicolumn{1}{c|}{1}             & 1           & \multicolumn{1}{c|}{4}                  & \multicolumn{1}{c|}{8}            & \multicolumn{1}{c|}{5}             & \multicolumn{1}{c|}{5}             & 6           \\ \hline
26           & radixsort                         & \multicolumn{1}{c|}{1}                  & \multicolumn{1}{c|}{3}            & \multicolumn{1}{c|}{2}             & \multicolumn{1}{c|}{2}             & 6           & \multicolumn{1}{c|}{3}                  & \multicolumn{1}{c|}{5}            & \multicolumn{1}{c|}{3}             & \multicolumn{1}{c|}{4}             & 14          \\ \hline
27           & recursive-activity-selector       & \multicolumn{1}{c|}{1.5}                & \multicolumn{1}{c|}{1.5}          & \multicolumn{1}{c|}{1.5}           & \multicolumn{1}{c|}{1.5}           & 1.5         & \multicolumn{1}{c|}{1}                  & \multicolumn{1}{c|}{1}            & \multicolumn{1}{c|}{1}             & \multicolumn{1}{c|}{2}             & 8           \\ \hline
28           & slow\_all\_pair\_shortest\_pair   & \multicolumn{1}{c|}{4.5}                & \multicolumn{1}{c|}{7.25}         & \multicolumn{1}{c|}{9}             & \multicolumn{1}{c|}{4.5}           & 7.25        & \multicolumn{1}{c|}{8}                  & \multicolumn{1}{c|}{11}           & \multicolumn{1}{c|}{13}            & \multicolumn{1}{c|}{8}             & 11          \\ \hline
29           & matrix\_multiplication            & \multicolumn{1}{c|}{1.33}               & \multicolumn{1}{c|}{2.33}         & \multicolumn{1}{c|}{2}             & \multicolumn{1}{c|}{1.33}          & 1.33        & \multicolumn{1}{c|}{3}                  & \multicolumn{1}{c|}{5}            & \multicolumn{1}{c|}{8}             & \multicolumn{1}{c|}{6}             & 13          \\ \hline
             & \textbf{MRR}                               & \multicolumn{1}{c|}{\textbf{0.819}}              & \multicolumn{1}{c|}{0.307}        & \multicolumn{1}{c|}{0.346}         & \multicolumn{1}{c|}{0.65}          & 0.517       & \multicolumn{1}{c|}{\textbf{0.606}}              & \multicolumn{1}{c|}{0.232}        & \multicolumn{1}{c|}{0.263}         & \multicolumn{1}{c|}{0.432}         & 0.256       \\ \hline
             
\end{tabular}
}
\end{table}

\begin{table}[ht]

\centering
\caption{The F-rank results for experiments on {\it Stony Brook code database} for the setting of C language. For {\it one project}, the results show the mean over all the ground truth projects implementing the algorithm. The best performing MRR result is shown in bold.}
\label{c}
\resizebox{.96\textwidth}{!}{
\begin{tabular}{c||c||ccccc||ccccc}
\hline
             & \textbf{}                   & \multicolumn{5}{c||}{\textbf{one project}}                                                                                                                                              & \multicolumn{5}{c}{\textbf{multiple projects}}                                                                                                                                        \\ \hline
\textbf{No.} & \textbf{Query}              & \multicolumn{1}{c|}{\textbf{Beryllium}} & \multicolumn{1}{c|}{\textbf{DCS}}      & \multicolumn{1}{c|}{\textbf{OCoR}}     & \multicolumn{1}{c|}{\textbf{CB+G}}     & \textbf{CB}       & \multicolumn{1}{c|}{\textbf{Beryllium}} & \multicolumn{1}{c|}{\textbf{DCS}}      & \multicolumn{1}{c|}{\textbf{OCoR}}     & \multicolumn{1}{c|}{\textbf{CB+G}}     & \textbf{CB}       \\ \hline
1            & any-segments-intersect      & \multicolumn{1}{c|}{1}                  & \multicolumn{1}{c|}{3}                 & \multicolumn{1}{c|}{2}                 & \multicolumn{1}{c|}{1}                 & 1                 & \multicolumn{1}{c|}{16}                 & \multicolumn{1}{c|}{18}                & \multicolumn{1}{c|}{29}                & \multicolumn{1}{c|}{16}                & 16                \\ \hline
2            & approx-vertex-cover         & \multicolumn{1}{c|}{1}                  & \multicolumn{1}{c|}{1}                 & \multicolumn{1}{c|}{3}                 & \multicolumn{1}{c|}{5}                 & 32                & \multicolumn{1}{c|}{1}                  & \multicolumn{1}{c|}{1}                 & \multicolumn{1}{c|}{3}                 & \multicolumn{1}{c|}{6}                 & 41                \\ \hline
3            & breadth-first-search        & \multicolumn{1}{c|}{14}                 & \multicolumn{1}{c|}{6}                 & \multicolumn{1}{c|}{15}                & \multicolumn{1}{c|}{17}                & 32                & \multicolumn{1}{c|}{23}                 & \multicolumn{1}{c|}{8}                 & \multicolumn{1}{c|}{19}                & \multicolumn{1}{c|}{28}                & \textgreater{}100 \\ \hline
4            & extend-shortest-paths       & \multicolumn{1}{c|}{1.33}               & \multicolumn{1}{c|}{3}                 & \multicolumn{1}{c|}{2.33}              & \multicolumn{1}{c|}{1.33}              & 1.33              & \multicolumn{1}{c|}{1}                  & \multicolumn{1}{c|}{4}                 & \multicolumn{1}{c|}{2}                 & \multicolumn{1}{c|}{1}                 & 3                 \\ \hline
5            & graham-scan                 & \multicolumn{1}{c|}{19}                 & \multicolumn{1}{c|}{15.5}              & \multicolumn{1}{c|}{10.5}              & \multicolumn{1}{c|}{19.5}              & 50                & \multicolumn{1}{c|}{\textgreater{}100}  & \multicolumn{1}{c|}{15}                & \multicolumn{1}{c|}{15}                & \multicolumn{1}{c|}{\textgreater{}100} & \textgreater{}100 \\ \hline
6            & hopcroft-karp               & \multicolumn{1}{c|}{3}                  & \multicolumn{1}{c|}{1}                 & \multicolumn{1}{c|}{10}                & \multicolumn{1}{c|}{7}                 & 11                & \multicolumn{1}{c|}{31}                 & \multicolumn{1}{c|}{10}                & \multicolumn{1}{c|}{41}                & \multicolumn{1}{c|}{33}                & \textgreater{}100 \\ \hline
7            & insertionsort               & \multicolumn{1}{c|}{1.5}                & \multicolumn{1}{c|}{1.5}               & \multicolumn{1}{c|}{1.5}               & \multicolumn{1}{c|}{1.5}               & 1.5               & \multicolumn{1}{c|}{1}                  & \multicolumn{1}{c|}{19}                & \multicolumn{1}{c|}{24}                & \multicolumn{1}{c|}{1}                 & 1                 \\ \hline
8            & kruskal                     & \multicolumn{1}{c|}{6.33}               & \multicolumn{1}{c|}{7}                 & \multicolumn{1}{c|}{7.33}              & \multicolumn{1}{c|}{6.33}              & 6.33              & \multicolumn{1}{c|}{40}                 & \multicolumn{1}{c|}{39}                & \multicolumn{1}{c|}{\textgreater{}100} & \multicolumn{1}{c|}{40}                & \textgreater{}100 \\ \hline
9            & lcs-length                  & \multicolumn{1}{c|}{1}                  & \multicolumn{1}{c|}{1}                 & \multicolumn{1}{c|}{1}                 & \multicolumn{1}{c|}{1}                 & 1                 & \multicolumn{1}{c|}{1}                  & \multicolumn{1}{c|}{1}                 & \multicolumn{1}{c|}{1}                 & \multicolumn{1}{c|}{1}                 & 1                 \\ \hline
10           & maybe-mst                   & \multicolumn{1}{c|}{7.2}                & \multicolumn{1}{c|}{14.4}              & \multicolumn{1}{c|}{16.2}              & \multicolumn{1}{c|}{8.2}               & 8.6               & \multicolumn{1}{c|}{17}                 & \multicolumn{1}{c|}{21}                & \multicolumn{1}{c|}{26}                & \multicolumn{1}{c|}{19}                & \textgreater{}100 \\ \hline
11           & mst                         & \multicolumn{1}{c|}{1.4}                & \multicolumn{1}{c|}{7.2}               & \multicolumn{1}{c|}{8}                 & \multicolumn{1}{c|}{1.4}                 & 2.2               & \multicolumn{1}{c|}{7}                  & \multicolumn{1}{c|}{10}                & \multicolumn{1}{c|}{14}                & \multicolumn{1}{c|}{9}                 & 9                 \\ \hline
12           & mst-reduce                  & \multicolumn{1}{c|}{11.6}               & \multicolumn{1}{c|}{4.4}               & \multicolumn{1}{c|}{2.2}               & \multicolumn{1}{c|}{12}                & 18.8              & \multicolumn{1}{c|}{16}                 & \multicolumn{1}{c|}{12}                & \multicolumn{1}{c|}{9}                 & \multicolumn{1}{c|}{18}                & 20                \\ \hline
13           & optimal-bst                 & \multicolumn{1}{c|}{1}                  & \multicolumn{1}{c|}{4}                 & \multicolumn{1}{c|}{6}                 & \multicolumn{1}{c|}{2}                 & 2                 & \multicolumn{1}{c|}{11}                 & \multicolumn{1}{c|}{14}                & \multicolumn{1}{c|}{14}                & \multicolumn{1}{c|}{12}                & 13                \\ \hline
14           & prim                        & \multicolumn{1}{c|}{2}                  & \multicolumn{1}{c|}{2.8}               & \multicolumn{1}{c|}{2.2}               & \multicolumn{1}{c|}{2.2}               & 2.4               & \multicolumn{1}{c|}{49}                 & \multicolumn{1}{c|}{48}                & \multicolumn{1}{c|}{41}                & \multicolumn{1}{c|}{49}                & \textgreater{}100 \\ \hline
15           & quicksort                   & \multicolumn{1}{c|}{\textgreater{}100}  & \multicolumn{1}{c|}{\textgreater{}100} & \multicolumn{1}{c|}{\textgreater{}100} & \multicolumn{1}{c|}{\textgreater{}100} & \textgreater{}100 & \multicolumn{1}{c|}{\textgreater{}100}  & \multicolumn{1}{c|}{\textgreater{}100} & \multicolumn{1}{c|}{\textgreater{}100} & \multicolumn{1}{c|}{\textgreater{}100} & \textgreater{}100 \\ \hline
16           & radixsort                   & \multicolumn{1}{c|}{1.5}                & \multicolumn{1}{c|}{2.5}               & \multicolumn{1}{c|}{4}                 & \multicolumn{1}{c|}{1.5}               & 1.5               & \multicolumn{1}{c|}{1}                  & \multicolumn{1}{c|}{3}                 & \multicolumn{1}{c|}{5}                 & \multicolumn{1}{c|}{1}                 & 2                 \\ \hline
17           & recursive-activity-selector & \multicolumn{1}{c|}{1}                  & \multicolumn{1}{c|}{11}                & \multicolumn{1}{c|}{13.5}              & \multicolumn{1}{c|}{20}                & 21.5              & \multicolumn{1}{c|}{31}                 & \multicolumn{1}{c|}{43}                & \multicolumn{1}{c|}{47}                & \multicolumn{1}{c|}{35}                & \textgreater{}100 \\ \hline
18           & matrix\_multiplication      & \multicolumn{1}{c|}{4}                  & \multicolumn{1}{c|}{6}                 & \multicolumn{1}{c|}{7}                 & \multicolumn{1}{c|}{4}                 & 9                 & \multicolumn{1}{c|}{10}                 & \multicolumn{1}{c|}{12}                & \multicolumn{1}{c|}{15}                & \multicolumn{1}{c|}{12}                & 14                \\ \hline
             & \textbf{MRR}                         & \multicolumn{1}{c|}{\textbf{0.522}}              & \multicolumn{1}{c|}{0.357}             & \multicolumn{1}{c|}{0.281}             & \multicolumn{1}{c|}{0.472}             & 0.34              & \multicolumn{1}{c|}{\textbf{0.316}}              & \multicolumn{1}{c|}{0.192}             & \multicolumn{1}{c|}{0.149}             & \multicolumn{1}{c|}{0.265}             & 0.184             \\ \hline
\end{tabular}
}
\end{table}

\begin{table}[ht]

\centering
\caption{The F-rank results for experiments on {\it Stony Brook code database} for the setting of mixed language. For {\it one project}, the results show the mean over all the ground truth projects implementing the algorithm. The best performing MRR result is shown in bold.}
\label{mixed}
\resizebox{.96\textwidth}{!}{
\begin{tabular}{c||c||ccccc||ccccc}
\hline
             & \textbf{}                   & \multicolumn{5}{c||}{\textbf{one project}}                                                                                                                           & \multicolumn{5}{c}{\textbf{multiple projects}}                                                                                                    \\ \hline
\textbf{No.} & \textbf{Query}              & \multicolumn{1}{c|}{\textbf{Beryllium}} & \multicolumn{1}{c|}{\textbf{DCS}} & \multicolumn{1}{c|}{\textbf{OCoR}} & \multicolumn{1}{c|}{\textbf{CB+G}} & \textbf{CB} & \multicolumn{1}{c|}{\textbf{Beryllium}} & \multicolumn{1}{c|}{\textbf{DCS}} & \multicolumn{1}{c|}{\textbf{OCoR}} & \multicolumn{1}{c|}{\textbf{CB+G}} & \textbf{CB} \\ \hline
1            & any-segments-intersect      & \multicolumn{1}{c|}{10}                 & \multicolumn{1}{c|}{12.33}        & \multicolumn{1}{c|}{16}            & \multicolumn{1}{c|}{10.33}         & 12.33       & \multicolumn{1}{c|}{12}                 & \multicolumn{1}{c|}{16}           & \multicolumn{1}{c|}{19}            & \multicolumn{1}{c|}{14}            & 18          \\ \hline
2            & approx-vertex-cover         & \multicolumn{1}{c|}{1.5}                & \multicolumn{1}{c|}{1.5}          & \multicolumn{1}{c|}{3}             & \multicolumn{1}{c|}{1.5}           & 2.5         & \multicolumn{1}{c|}{1}                  & \multicolumn{1}{c|}{3}            & \multicolumn{1}{c|}{4}             & \multicolumn{1}{c|}{2}             & 5           \\ \hline
3            & breadth-first-search        & \multicolumn{1}{c|}{6}                  & \multicolumn{1}{c|}{6.2}          & \multicolumn{1}{c|}{4.2}           & \multicolumn{1}{c|}{6.2}           & 9.4         & \multicolumn{1}{c|}{12}                 & \multicolumn{1}{c|}{13}           & \multicolumn{1}{c|}{6}             & \multicolumn{1}{c|}{13}            & 16          \\ \hline
4            & extend-shortest-paths       & \multicolumn{1}{c|}{1.33}               & \multicolumn{1}{c|}{4.33}         & \multicolumn{1}{c|}{2.16}          & \multicolumn{1}{c|}{3.16}          & 5.33        & \multicolumn{1}{c|}{1}                  & \multicolumn{1}{c|}{5}            & \multicolumn{1}{c|}{2}             & \multicolumn{1}{c|}{3}             & 5           \\ \hline
5            & graham-scan                 & \multicolumn{1}{c|}{9}                  & \multicolumn{1}{c|}{7.25}         & \multicolumn{1}{c|}{9.25}          & \multicolumn{1}{c|}{9.25}          & 10.25       & \multicolumn{1}{c|}{14}                 & \multicolumn{1}{c|}{12}           & \multicolumn{1}{c|}{16}            & \multicolumn{1}{c|}{14}            & 16          \\ \hline
6            & hopcroft-karp               & \multicolumn{1}{c|}{3.33}               & \multicolumn{1}{c|}{1.33}         & \multicolumn{1}{c|}{17}            & \multicolumn{1}{c|}{3.33}          & 3.33        & \multicolumn{1}{c|}{15}                 & \multicolumn{1}{c|}{5}            & \multicolumn{1}{c|}{29}            & \multicolumn{1}{c|}{15}            & 19          \\ \hline
7            & insertionsort               & \multicolumn{1}{c|}{1.33}               & \multicolumn{1}{c|}{2}            & \multicolumn{1}{c|}{3.33}          & \multicolumn{1}{c|}{3}             & 4.33        & \multicolumn{1}{c|}{2}                  & \multicolumn{1}{c|}{4}            & \multicolumn{1}{c|}{3}             & \multicolumn{1}{c|}{3}             & 7           \\ \hline
8            & kruskal                     & \multicolumn{1}{c|}{1.28}               & \multicolumn{1}{c|}{6.42}         & \multicolumn{1}{c|}{7.14}          & \multicolumn{1}{c|}{2.14}          & 3           & \multicolumn{1}{c|}{2}                  & \multicolumn{1}{c|}{13}           & \multicolumn{1}{c|}{12}            & \multicolumn{1}{c|}{2}             & 3           \\ \hline
9            & lcs-length                  & \multicolumn{1}{c|}{1.5}                & \multicolumn{1}{c|}{3.5}          & \multicolumn{1}{c|}{1.5}           & \multicolumn{1}{c|}{3.5}           & 5           & \multicolumn{1}{c|}{1}                  & \multicolumn{1}{c|}{3}            & \multicolumn{1}{c|}{1}             & \multicolumn{1}{c|}{3}             & 4           \\ \hline
10           & maybe-mst                   & \multicolumn{1}{c|}{1.3}                & \multicolumn{1}{c|}{12.5}         & \multicolumn{1}{c|}{14.4}          & \multicolumn{1}{c|}{1.4}           & 2.3         & \multicolumn{1}{c|}{10}                 & \multicolumn{1}{c|}{15}           & \multicolumn{1}{c|}{17}            & \multicolumn{1}{c|}{11}            & 13          \\ \hline
11           & mst                         & \multicolumn{1}{c|}{1.4}                & \multicolumn{1}{c|}{2.3}          & \multicolumn{1}{c|}{1.4}           & \multicolumn{1}{c|}{1.4}           & 3.2         & \multicolumn{1}{c|}{3}                  & \multicolumn{1}{c|}{3}            & \multicolumn{1}{c|}{2}             & \multicolumn{1}{c|}{4}             & 7           \\ \hline
12           & mst-reduce                  & \multicolumn{1}{c|}{8.1}                & \multicolumn{1}{c|}{9.2}          & \multicolumn{1}{c|}{9.3}           & \multicolumn{1}{c|}{8.4}           & 11.2        & \multicolumn{1}{c|}{9}                  & \multicolumn{1}{c|}{9}            & \multicolumn{1}{c|}{9}             & \multicolumn{1}{c|}{9}             & 13          \\ \hline
13           & optimal-bst                 & \multicolumn{1}{c|}{2.2}                & \multicolumn{1}{c|}{6.4}          & \multicolumn{1}{c|}{8.2}           & \multicolumn{1}{c|}{3.4}           & 4.2         & \multicolumn{1}{c|}{6}                  & \multicolumn{1}{c|}{12}           & \multicolumn{1}{c|}{14}            & \multicolumn{1}{c|}{6}             & 7           \\ \hline
14           & prim                        & \multicolumn{1}{c|}{2.22}               & \multicolumn{1}{c|}{3}            & \multicolumn{1}{c|}{2.22}          & \multicolumn{1}{c|}{3.22}          & 3.33        & \multicolumn{1}{c|}{14}                 & \multicolumn{1}{c|}{21}           & \multicolumn{1}{c|}{19}            & \multicolumn{1}{c|}{14}            & 14          \\ \hline
15           & quicksort                   & \multicolumn{1}{c|}{10.33}              & \multicolumn{1}{c|}{13.33}        & \multicolumn{1}{c|}{12}            & \multicolumn{1}{c|}{12.33}         & 15          & \multicolumn{1}{c|}{56}                 & \multicolumn{1}{c|}{59}           & \multicolumn{1}{c|}{57}            & \multicolumn{1}{c|}{57}            & 71          \\ \hline
16           & radixsort                   & \multicolumn{1}{c|}{1.33}               & \multicolumn{1}{c|}{1.33}         & \multicolumn{1}{c|}{1.33}          & \multicolumn{1}{c|}{2.33}          & 5           & \multicolumn{1}{c|}{9}                  & \multicolumn{1}{c|}{7}            & \multicolumn{1}{c|}{5}             & \multicolumn{1}{c|}{9}             & 9           \\ \hline
17           & recursive-activity-selector & \multicolumn{1}{c|}{1.25}               & \multicolumn{1}{c|}{9.25}         & \multicolumn{1}{c|}{11}            & \multicolumn{1}{c|}{2.25}          & 2.5         & \multicolumn{1}{c|}{4}                  & \multicolumn{1}{c|}{13}           & \multicolumn{1}{c|}{17}            & \multicolumn{1}{c|}{6}             & 9           \\ \hline
18           & matrix\_multiplication      & \multicolumn{1}{c|}{2}                  & \multicolumn{1}{c|}{2.25}         & \multicolumn{1}{c|}{2.25}          & \multicolumn{1}{c|}{2.25}          & 2.5         & \multicolumn{1}{c|}{6}                  & \multicolumn{1}{c|}{8}            & \multicolumn{1}{c|}{10}            & \multicolumn{1}{c|}{8}             & 11          \\ \hline
             & \textbf{MRR}                         & \multicolumn{1}{c|}{\textbf{0.498}}              & \multicolumn{1}{c|}{0.304}        & \multicolumn{1}{c|}{0.289}         & \multicolumn{1}{c|}{0.349}         & 0.243       & \multicolumn{1}{c|}{\textbf{0.313}}              & \multicolumn{1}{c|}{0.146}        & \multicolumn{1}{c|}{0.203}         & \multicolumn{1}{c|}{0.189}         & 0.122       \\ \hline
\end{tabular}
}
\end{table}

\begin{table}[ht]

\centering
\caption{The F-rank results for experiments on {\it GitHub code database} for the setting of Java language. The best performing MRR result is shown in bold.}
\label{github java}
\resizebox{.68\textwidth}{!}{
\begin{tabular}{c||c||ccccc}
\hline
             & \textbf{}                         & \multicolumn{5}{c}{\textbf{GitHub projects (Java)}}                                                                                                                               \\ \hline
\textbf{No.} & \textbf{Query}                    & \multicolumn{1}{c|}{\textbf{Beryllium}} & \multicolumn{1}{c|}{\textbf{DCS}}      & \multicolumn{1}{c|}{\textbf{OCoR}}     & \multicolumn{1}{c|}{\textbf{CB+G}}     & \textbf{CB}       \\ \hline
1            & any-segments-intersect            & \multicolumn{1}{c|}{\textgreater{}100}  & \multicolumn{1}{c|}{\textgreater{}100} & \multicolumn{1}{c|}{\textgreater{}100} & \multicolumn{1}{c|}{\textgreater{}100} & \textgreater{}100 \\ \hline
2            & approx-vertex-cover               & \multicolumn{1}{c|}{1}                  & \multicolumn{1}{c|}{5}                 & \multicolumn{1}{c|}{3}                 & \multicolumn{1}{c|}{2}                 & 4                 \\ \hline
3            & breadth-first-search              & \multicolumn{1}{c|}{1}                  & \multicolumn{1}{c|}{7}                 & \multicolumn{1}{c|}{5}                 & \multicolumn{1}{c|}{1}                 & 2                 \\ \hline
4            & compute-transition-function       & \multicolumn{1}{c|}{2}                  & \multicolumn{1}{c|}{4}                 & \multicolumn{1}{c|}{5}                 & \multicolumn{1}{c|}{4}                 & 13                \\ \hline
5            & extend-shortest-paths             & \multicolumn{1}{c|}{4}                  & \multicolumn{1}{c|}{21}                & \multicolumn{1}{c|}{12}                & \multicolumn{1}{c|}{5}                 & 10                \\ \hline
6            & finite-automaton-matcher          & \multicolumn{1}{c|}{1}                  & \multicolumn{1}{c|}{9}                 & \multicolumn{1}{c|}{13}                & \multicolumn{1}{c|}{2}                 & 73                \\ \hline
7            & floyd-warshall                    & \multicolumn{1}{c|}{8}                  & \multicolumn{1}{c|}{10}                & \multicolumn{1}{c|}{9}                 & \multicolumn{1}{c|}{8}                 & 12                \\ \hline
8            & graham-scan                       & \multicolumn{1}{c|}{1}                  & \multicolumn{1}{c|}{9}                 & \multicolumn{1}{c|}{5}                 & \multicolumn{1}{c|}{1}                 & 4                 \\ \hline
9            & hopcroft-karp                     & \multicolumn{1}{c|}{1}                  & \multicolumn{1}{c|}{7}                 & \multicolumn{1}{c|}{5}                 & \multicolumn{1}{c|}{3}                 & 8                 \\ \hline
10           & insertionsort                     & \multicolumn{1}{c|}{1}                  & \multicolumn{1}{c|}{2}                 & \multicolumn{1}{c|}{3}                 & \multicolumn{1}{c|}{1}                 & 2                 \\ \hline
11           & johnson                           & \multicolumn{1}{c|}{\textgreater{}100}  & \multicolumn{1}{c|}{\textgreater{}100} & \multicolumn{1}{c|}{\textgreater{}100} & \multicolumn{1}{c|}{\textgreater{}100} & \textgreater{}100 \\ \hline
12           & kruskal                           & \multicolumn{1}{c|}{9}                  & \multicolumn{1}{c|}{15}                & \multicolumn{1}{c|}{12}                & \multicolumn{1}{c|}{10}                & 11                \\ \hline
13           & lcs-length                        & \multicolumn{1}{c|}{1}                  & \multicolumn{1}{c|}{16}                & \multicolumn{1}{c|}{7}                 & \multicolumn{1}{c|}{1}                 & 2                 \\ \hline
14           & lu-decomposition                  & \multicolumn{1}{c|}{10}                 & \multicolumn{1}{c|}{9}                 & \multicolumn{1}{c|}{7}                 & \multicolumn{1}{c|}{12}                & \textgreater{}100 \\ \hline
15           & maybe-mst                         & \multicolumn{1}{c|}{1}                  & \multicolumn{1}{c|}{8}                 & \multicolumn{1}{c|}{7}                 & \multicolumn{1}{c|}{1}                 & 2                 \\ \hline
16           & modular-linear-equation-solver    & \multicolumn{1}{c|}{13}                 & \multicolumn{1}{c|}{18}                & \multicolumn{1}{c|}{21}                & \multicolumn{1}{c|}{13}                & \textgreater{}100 \\ \hline
17           & mst                               & \multicolumn{1}{c|}{1}                  & \multicolumn{1}{c|}{16}                & \multicolumn{1}{c|}{12}                & \multicolumn{1}{c|}{2}                 & 4                 \\ \hline
18           & mst-reduce                        & \multicolumn{1}{c|}{\textgreater{}100}  & \multicolumn{1}{c|}{\textgreater{}100} & \multicolumn{1}{c|}{\textgreater{}100} & \multicolumn{1}{c|}{\textgreater{}100} & \textgreater{}100 \\ \hline
19           & naïve-string-matcher              & \multicolumn{1}{c|}{1}                  & \multicolumn{1}{c|}{11}                & \multicolumn{1}{c|}{21}                & \multicolumn{1}{c|}{2}                 & 3                 \\ \hline
20           & optimal-bst                       & \multicolumn{1}{c|}{\textgreater{}100}  & \multicolumn{1}{c|}{\textgreater{}100} & \multicolumn{1}{c|}{\textgreater{}100} & \multicolumn{1}{c|}{\textgreater{}100} & \textgreater{}100 \\ \hline
21           & pivot                             & \multicolumn{1}{c|}{\textgreater{}100}  & \multicolumn{1}{c|}{\textgreater{}100} & \multicolumn{1}{c|}{\textgreater{}100} & \multicolumn{1}{c|}{\textgreater{}100} & \textgreater{}100 \\ \hline
22           & prim                              & \multicolumn{1}{c|}{1}                  & \multicolumn{1}{c|}{6}                 & \multicolumn{1}{c|}{6}                 & \multicolumn{1}{c|}{4}                 & 9                 \\ \hline
23           & print\_all\_pair\_shortest\_paths & \multicolumn{1}{c|}{1}                  & \multicolumn{1}{c|}{6}                 & \multicolumn{1}{c|}{11}                & \multicolumn{1}{c|}{1}                 & 1                 \\ \hline
24           & quicksort                         & \multicolumn{1}{c|}{1}                  & \multicolumn{1}{c|}{11}                & \multicolumn{1}{c|}{16}                & \multicolumn{1}{c|}{2}                 & 4                 \\ \hline
25           & rabin-karp-matcher                & \multicolumn{1}{c|}{1}                  & \multicolumn{1}{c|}{3}                 & \multicolumn{1}{c|}{2}                 & \multicolumn{1}{c|}{1}                 & 2                 \\ \hline
26           & radixsort                         & \multicolumn{1}{c|}{16}                 & \multicolumn{1}{c|}{21}                & \multicolumn{1}{c|}{17}                & \multicolumn{1}{c|}{20}                & 60                \\ \hline
27           & recursive-activity-selector       & \multicolumn{1}{c|}{3}                  & \multicolumn{1}{c|}{12}                & \multicolumn{1}{c|}{5}                 & \multicolumn{1}{c|}{5}                 & 11                \\ \hline
28           & slow\_all\_pair\_shortest\_pair   & \multicolumn{1}{c|}{1}                  & \multicolumn{1}{c|}{14}                & \multicolumn{1}{c|}{9}                 & \multicolumn{1}{c|}{1}                 & 2                 \\ \hline
29           & matrix\_multiplication            & \multicolumn{1}{c|}{5}                  & \multicolumn{1}{c|}{10}                & \multicolumn{1}{c|}{6}                 & \multicolumn{1}{c|}{6}                 & 6                 \\ \hline
             & \textbf{MRR}                               & \multicolumn{1}{c|}{\textbf{0.58}}               & \multicolumn{1}{c|}{0.113}             & \multicolumn{1}{c|}{0.132}             & \multicolumn{1}{c|}{0.427}             & 0.217             \\ \hline
\end{tabular}
}
\end{table}
\begin{table}[ht]
\centering
\caption{The F-rank results for experiments on {\it GitHub code database} for the setting of C and mixed languages. The best performing MRR result is shown in bold.}
\label{github c and mixed}
\resizebox{.96\textwidth}{!}{
\begin{tabular}{c||c||ccccc||ccccc}
\hline
             & \textbf{}                         & \multicolumn{5}{c||}{\textbf{GitHub projects   (C)}}                                                                                                                                & \multicolumn{5}{c}{\textbf{GitHub projects   (mixed)}}                                                                                                                            \\ \hline
\textbf{No.} & \textbf{Query}                    & \multicolumn{1}{c|}{\textbf{Beryllium}} & \multicolumn{1}{c|}{\textbf{DCS}}      & \multicolumn{1}{c|}{\textbf{OCoR}}     & \multicolumn{1}{c|}{\textbf{CB+G}}     & \textbf{CB}       & \multicolumn{1}{c|}{\textbf{Beryllium}} & \multicolumn{1}{c|}{\textbf{DCS}}      & \multicolumn{1}{c|}{\textbf{OCoR}}     & \multicolumn{1}{c|}{\textbf{CB+G}}     & \textbf{CB}       \\ \hline
1            & any-segments-intersect            & \multicolumn{1}{c|}{\textgreater{}100}  & \multicolumn{1}{c|}{\textgreater{}100} & \multicolumn{1}{c|}{\textgreater{}100} & \multicolumn{1}{c|}{\textgreater{}100} & \textgreater{}100 & \multicolumn{1}{c|}{\textgreater{}100}  & \multicolumn{1}{c|}{\textgreater{}100} & \multicolumn{1}{c|}{\textgreater{}100} & \multicolumn{1}{c|}{\textgreater{}100} & \textgreater{}100 \\ \hline
2            & approx-vertex-cover               & \multicolumn{1}{c|}{10}                 & \multicolumn{1}{c|}{13}                & \multicolumn{1}{c|}{13}                & \multicolumn{1}{c|}{12}                & 18                & \multicolumn{1}{c|}{3}                  & \multicolumn{1}{c|}{8}                 & \multicolumn{1}{c|}{5}                 & \multicolumn{1}{c|}{9}                 & 12                \\ \hline
3            & breadth-first-search              & \multicolumn{1}{c|}{3}                  & \multicolumn{1}{c|}{\textgreater{}100} & \multicolumn{1}{c|}{2}                 & \multicolumn{1}{c|}{5}                 & 16                & \multicolumn{1}{c|}{2}                  & \multicolumn{1}{c|}{34}                & \multicolumn{1}{c|}{47}                & \multicolumn{1}{c|}{3}                 & 15                \\ \hline
4            & compute-transition-function       & \multicolumn{1}{c|}{10}                 & \multicolumn{1}{c|}{18}                & \multicolumn{1}{c|}{21}                & \multicolumn{1}{c|}{10}                & 19                & \multicolumn{1}{c|}{4}                  & \multicolumn{1}{c|}{24}                & \multicolumn{1}{c|}{25}                & \multicolumn{1}{c|}{6}                 & 15                \\ \hline
5            & extend-shortest-paths             & \multicolumn{1}{c|}{1}                  & \multicolumn{1}{c|}{4}                 & \multicolumn{1}{c|}{4}                 & \multicolumn{1}{c|}{1}                 & 2                 & \multicolumn{1}{c|}{2}                  & \multicolumn{1}{c|}{19}                & \multicolumn{1}{c|}{9}                 & \multicolumn{1}{c|}{2}                 & 3                 \\ \hline
6            & finite-automaton-matcher          & \multicolumn{1}{c|}{2}                  & \multicolumn{1}{c|}{12}                & \multicolumn{1}{c|}{8}                 & \multicolumn{1}{c|}{3}                 & 6                 & \multicolumn{1}{c|}{2}                  & \multicolumn{1}{c|}{10}                & \multicolumn{1}{c|}{11}                & \multicolumn{1}{c|}{4}                 & 16                \\ \hline
7            & floyd-warshall                    & \multicolumn{1}{c|}{4}                  & \multicolumn{1}{c|}{14}                & \multicolumn{1}{c|}{12}                & \multicolumn{1}{c|}{6}                 & 13                & \multicolumn{1}{c|}{6}                  & \multicolumn{1}{c|}{22}                & \multicolumn{1}{c|}{20}                & \multicolumn{1}{c|}{10}                & 18                \\ \hline
8            & graham-scan                       & \multicolumn{1}{c|}{5}                  & \multicolumn{1}{c|}{13}                & \multicolumn{1}{c|}{7}                 & \multicolumn{1}{c|}{8}                 & 28                & \multicolumn{1}{c|}{4}                  & \multicolumn{1}{c|}{10}                & \multicolumn{1}{c|}{6}                 & \multicolumn{1}{c|}{6}                 & 26                \\ \hline
9            & hopcroft-karp                     & \multicolumn{1}{c|}{1}                  & \multicolumn{1}{c|}{13}                & \multicolumn{1}{c|}{12}                & \multicolumn{1}{c|}{1}                 & 3                 & \multicolumn{1}{c|}{1}                  & \multicolumn{1}{c|}{15}                & \multicolumn{1}{c|}{18}                & \multicolumn{1}{c|}{4}                 & 8                 \\ \hline
10           & insertionsort                     & \multicolumn{1}{c|}{6}                  & \multicolumn{1}{c|}{1}                 & \multicolumn{1}{c|}{2}                 & \multicolumn{1}{c|}{8}                 & 20                & \multicolumn{1}{c|}{4}                  & \multicolumn{1}{c|}{4}                 & \multicolumn{1}{c|}{5}                 & \multicolumn{1}{c|}{5}                 & 16                \\ \hline
11           & johnson                           & \multicolumn{1}{c|}{18}                 & \multicolumn{1}{c|}{\textgreater{}100} & \multicolumn{1}{c|}{\textgreater{}100} & \multicolumn{1}{c|}{24}                & 32                & \multicolumn{1}{c|}{26}                 & \multicolumn{1}{c|}{\textgreater{}100} & \multicolumn{1}{c|}{\textgreater{}100} & \multicolumn{1}{c|}{42}                & 72                \\ \hline
12           & kruskal                           & \multicolumn{1}{c|}{41}                 & \multicolumn{1}{c|}{12}                & \multicolumn{1}{c|}{16}                & \multicolumn{1}{c|}{42}                & 48                & \multicolumn{1}{c|}{12}                 & \multicolumn{1}{c|}{14}                & \multicolumn{1}{c|}{13}                & \multicolumn{1}{c|}{15}                & 18                \\ \hline
13           & lcs-length                        & \multicolumn{1}{c|}{\textgreater{}100}  & \multicolumn{1}{c|}{\textgreater{}100} & \multicolumn{1}{c|}{4}                 & \multicolumn{1}{c|}{\textgreater{}100} & \textgreater{}100 & \multicolumn{1}{c|}{2}                  & \multicolumn{1}{c|}{62}                & \multicolumn{1}{c|}{8}                 & \multicolumn{1}{c|}{2}                 & 4                 \\ \hline
14           & lu-decomposition                  & \multicolumn{1}{c|}{8}                  & \multicolumn{1}{c|}{9}                 & \multicolumn{1}{c|}{8}                 & \multicolumn{1}{c|}{9}                 & 13                & \multicolumn{1}{c|}{9}                  & \multicolumn{1}{c|}{9}                 & \multicolumn{1}{c|}{8}                 & \multicolumn{1}{c|}{13}                & 18                \\ \hline
15           & maybe-mst                         & \multicolumn{1}{c|}{3}                  & \multicolumn{1}{c|}{14}                & \multicolumn{1}{c|}{10}                & \multicolumn{1}{c|}{4}                 & 20                & \multicolumn{1}{c|}{2}                  & \multicolumn{1}{c|}{16}                & \multicolumn{1}{c|}{12}                & \multicolumn{1}{c|}{4}                 & 15                \\ \hline
16           & modular-linear-equation-solver    & \multicolumn{1}{c|}{\textgreater{}100}  & \multicolumn{1}{c|}{\textgreater{}100} & \multicolumn{1}{c|}{\textgreater{}100} & \multicolumn{1}{c|}{\textgreater{}100} & \textgreater{}100 & \multicolumn{1}{c|}{15}                 & \multicolumn{1}{c|}{21}                & \multicolumn{1}{c|}{23}                & \multicolumn{1}{c|}{18}                & \textgreater{}100 \\ \hline
17           & mst                               & \multicolumn{1}{c|}{3}                  & \multicolumn{1}{c|}{12}                & \multicolumn{1}{c|}{9}                 & \multicolumn{1}{c|}{4}                 & 21                & \multicolumn{1}{c|}{2}                  & \multicolumn{1}{c|}{12}                & \multicolumn{1}{c|}{14}                & \multicolumn{1}{c|}{3}                 & 12                \\ \hline
18           & mst-reduce                        & \multicolumn{1}{c|}{\textgreater{}100}  & \multicolumn{1}{c|}{\textgreater{}100} & \multicolumn{1}{c|}{\textgreater{}100} & \multicolumn{1}{c|}{\textgreater{}100} & \textgreater{}100 & \multicolumn{1}{c|}{\textgreater{}100}  & \multicolumn{1}{c|}{\textgreater{}100} & \multicolumn{1}{c|}{\textgreater{}100} & \multicolumn{1}{c|}{\textgreater{}100} & \textgreater{}100 \\ \hline
19           & naïve-string-matcher              & \multicolumn{1}{c|}{2}                  & \multicolumn{1}{c|}{9}                 & \multicolumn{1}{c|}{6}                 & \multicolumn{1}{c|}{2}                 & 4                 & \multicolumn{1}{c|}{2}                  & \multicolumn{1}{c|}{11}                & \multicolumn{1}{c|}{12}                & \multicolumn{1}{c|}{4}                 & 5                 \\ \hline
20           & optimal-bst                       & \multicolumn{1}{c|}{\textgreater{}100}  & \multicolumn{1}{c|}{\textgreater{}100} & \multicolumn{1}{c|}{\textgreater{}100} & \multicolumn{1}{c|}{\textgreater{}100} & \textgreater{}100 & \multicolumn{1}{c|}{\textgreater{}100}  & \multicolumn{1}{c|}{\textgreater{}100} & \multicolumn{1}{c|}{\textgreater{}100} & \multicolumn{1}{c|}{\textgreater{}100} & \textgreater{}100 \\ \hline
21           & pivot                             & \multicolumn{1}{c|}{23}                 & \multicolumn{1}{c|}{38}                & \multicolumn{1}{c|}{42}                & \multicolumn{1}{c|}{26}                & \textgreater{}100 & \multicolumn{1}{c|}{36}                 & \multicolumn{1}{c|}{54}                & \multicolumn{1}{c|}{65}                & \multicolumn{1}{c|}{43}                & \textgreater{}100 \\ \hline
22           & prim                              & \multicolumn{1}{c|}{1}                  & \multicolumn{1}{c|}{7}                 & \multicolumn{1}{c|}{16}                & \multicolumn{1}{c|}{2}                 & 36                & \multicolumn{1}{c|}{1}                  & \multicolumn{1}{c|}{6}                 & \multicolumn{1}{c|}{9}                 & \multicolumn{1}{c|}{3}                 & 34                \\ \hline
23           & print\_all\_pair\_shortest\_paths & \multicolumn{1}{c|}{13}                 & \multicolumn{1}{c|}{10}                & \multicolumn{1}{c|}{12}                & \multicolumn{1}{c|}{17}                & 21                & \multicolumn{1}{c|}{4}                  & \multicolumn{1}{c|}{19}                & \multicolumn{1}{c|}{11}                & \multicolumn{1}{c|}{4}                 & 5                 \\ \hline
24           & quicksort                         & \multicolumn{1}{c|}{1}                  & \multicolumn{1}{c|}{11}                & \multicolumn{1}{c|}{9}                 & \multicolumn{1}{c|}{1}                 & 18                & \multicolumn{1}{c|}{1}                  & \multicolumn{1}{c|}{12}                & \multicolumn{1}{c|}{12}                & \multicolumn{1}{c|}{1}                 & 14                \\ \hline
25           & rabin-karp-matcher                & \multicolumn{1}{c|}{2}                  & \multicolumn{1}{c|}{3}                 & \multicolumn{1}{c|}{2}                 & \multicolumn{1}{c|}{3}                 & 5                 & \multicolumn{1}{c|}{2}                  & \multicolumn{1}{c|}{3}                 & \multicolumn{1}{c|}{2}                 & \multicolumn{1}{c|}{2}                 & 4                 \\ \hline
26           & radixsort                         & \multicolumn{1}{c|}{15}                 & \multicolumn{1}{c|}{24}                & \multicolumn{1}{c|}{26}                & \multicolumn{1}{c|}{18}                & 45                & \multicolumn{1}{c|}{18}                 & \multicolumn{1}{c|}{21}                & \multicolumn{1}{c|}{19}                & \multicolumn{1}{c|}{21}                & 52                \\ \hline
27           & recursive-activity-selector       & \multicolumn{1}{c|}{2}                  & \multicolumn{1}{c|}{9}                 & \multicolumn{1}{c|}{13}                & \multicolumn{1}{c|}{2}                 & 16                & \multicolumn{1}{c|}{3}                  & \multicolumn{1}{c|}{9}                 & \multicolumn{1}{c|}{7}                 & \multicolumn{1}{c|}{4}                 & 14                \\ \hline
28           & slow\_all\_pair\_shortest\_pair   & \multicolumn{1}{c|}{27}                 & \multicolumn{1}{c|}{24}                & \multicolumn{1}{c|}{21}                & \multicolumn{1}{c|}{31}                & \textgreater{}100 & \multicolumn{1}{c|}{4}                  & \multicolumn{1}{c|}{16}                & \multicolumn{1}{c|}{12}                & \multicolumn{1}{c|}{4}                 & 6                 \\ \hline
29           & matrix\_multiplication            & \multicolumn{1}{c|}{1}                  & \multicolumn{1}{c|}{6}                 & \multicolumn{1}{c|}{4}                 & \multicolumn{1}{c|}{3}                 & 28                & \multicolumn{1}{c|}{2}                  & \multicolumn{1}{c|}{8}                 & \multicolumn{1}{c|}{6}                 & \multicolumn{1}{c|}{3}                 & 26                \\ \hline
             & \textbf{MRR}                               & \multicolumn{1}{c|}{\textbf{0.321}}              & \multicolumn{1}{c|}{0.113}             & \multicolumn{1}{c|}{0.133}             & \multicolumn{1}{c|}{0.249}             & 0.08              & \multicolumn{1}{c|}{\textbf{0.345}}              & \multicolumn{1}{c|}{0.081}             & \multicolumn{1}{c|}{0.098}             & \multicolumn{1}{c|}{0.229}             & 0.087             \\ \hline
\end{tabular}
}
\end{table}

\textbf{Experiment 1 (one project):} In Tables~\ref{java} and \ref{c}, under {\it one project}, we observe that \toolname found the implementation for $29$ out of $29$ queries from Java projects and $17$ out of $18$ queries from C projects in {\it Stony Brook code database}. Among these, we found implementation for $25$ out of $29$ queries from Java projects and $9$ out of $18$ queries from C projects as the first hit. Table~\ref{mixed} shows that \toolname can have the {\it GAE} models trained based on the mixed languages of Java and C language and retrieve the implementations from both Java and C projects. In this setting, \toolname found the implementations for $9$ out of $18$ queries as the first hit. Overall, across all the settings in Tables~\ref{java}, \ref{c}, and \ref{mixed}, except for one query, \toolname found all the implementations of known queries in the top $25$ code snippets. In terms of MRR, \toolname performed the best on Java, followed by the mixed language model, followed by C. This is likely caused by the fact that CodeBERT is pre-trained on Java, so it can model Java features better.

\toolname outperforms the baselines {\it DCS} and {\it OCoR} for all the queries in Table~\ref{java}, for most queries in Table~\ref{c} except for queries No.3, No.5, No.6, and No.12, and for most queries in Table~\ref{mixed} except for queries No.5 and No.6. Among these queries, {\it DCS} performed the best for queries No.3 and No.6 in Table~\ref{c} and No.5 and No.6 queries in Table~\ref{mixed} while {\it OCoR} performed the best on the remaining.  In these cases, the value of the F-rank is lower for \toolname because \toolname has more candidate code fragments to rank----the baselines report methods and \toolname reports code fragments; in \toolname, one method can have several code fragments in the ranking. In return, \toolname can more accurately locate the algorithm implementations within a function or across multiple functions.

\toolname achieved significantly higher MRR than both baselines. Specifically, \toolname outperformed {\it DCS} by a margin of $167\%$ and {\it OCoR} by a margin of $137\%$ for the setting of Java language. For the setting of C language, \toolname outperformed {\it DCS} by a margin of $46\%$ and {\it OCoR} by a margin of $86\%$. For the setting of mixed language, \toolname outperformed {\it DCS} by a margin of $64\%$ and {\it OCoR} by a margin of $72\%$.

\textbf{Experiment 2 (multiple projects):}
In Tables~\ref{java}, \ref{c}, and~\ref{mixed}, under {\it multiple projects}, we can observe that within the top 25 of the ranked lists, \toolname found the implementation for $28$ out of $29$ queries from Java projects, $12$ out of $18$ queries from C projects, and $17$ out of $18$ queries from mixed language projects in Stony Brook code database. \toolname outperformed the baselines {\it DCS} and {\it OCoR} for $40$ queries out of the $65$ queries. {\it OCoR} won for query No.10 in Table~\ref{java} and No.11 and No.16 in Table~\ref{mixed}, and {\it DCS} performed the best for the remaining.  In terms of MRR, \toolname outperformed {\it DCS} by a margin of $161\%$ and {\it OCoR} by a margin of $130\%$ for the setting of Java language. For the setting of C language, \toolname outperformed {\it DCS} by a margin of $65\%$ and {\it OCoR} by a margin of $112\%$. For the setting of mixed language, \toolname outperformed {\it DCS} by a margin of $114\%$ and {\it OCoR} by a margin of $54\%$.

\subsubsection{GitHub code database}
 
In Tables~\ref{github java} and~\ref{github c and mixed}, under {\it GitHub projects}, we show that within the top $25$ of the ranked lists,  \toolname found the implementation for $24$ out of $29$ queries from Java projects, for $24$ out of $29$ queries from C projects, and for $26$ out of $29$ queries from mixed language projects. Excluding the $3$ queries that none of the methods retrieved implementations in top $100$ results, \toolname outperformed the baselines {\it DCS} and {\it OCoR} for $62$ of the $84$ queries. {\it OCoR} performed the best for $3$ out of $84$ queries, while {\it DCS} performed the best for $1$ out of $84$ queries. We observed that the library calls and complex data structure initialization could lead to the mismatch of pseudo code and source code in our approach. In terms of MRR, \toolname outperformed {\it DCS} by a margin of $413\%$ and {\it OCoR} by a margin of $339\%$ for the setting of Java language. For the setting of C language, \toolname outperformed {\it DCS} by a margin of $184\%$ and {\it OCoR} by a margin of $141\%$. For the setting of mixed language, \toolname outperformed {\it DCS} by a margin of $326\%$ and {\it OCoR} by a margin of $252\%$.
 
Overall, we observe that the baselines perform well when the queries and code are more similar in natural languages because the baselines are trained considering the queries to be natural language. However, \toolname considers the control flow, maths, and natural language features; thus, it is more robust and general for all types of pseudo code queries.

\subsubsection{Ablation results}~\label{sec:3}
Tables~\ref{java}, \ref{c}, \ref{mixed}, \ref{github java}, and \ref{github c and mixed} show that \toolname performs better than {\it CB+G} and {\it CB}. For example, under {\it one project}, for $36$ out of $65$ queries, \toolname outperformed {\it CB+G}, and for $45$ out of $65$ queries, {\it CB+G} is better than {\it CB}. The results demonstrated that control flow and math information is useful. We also found that for the setting of the Java language, {\it CB} can find algorithm implementations better than our baselines. For the setting of the C language, CB performed poorly compared to the baselines, as CB uses encoding from CodeBERT, and CodeBERT is not trained in the C language. We observed that {\it CB+G} is significantly better than {\it CB} on the setting of C language, indicating that the control flow is very useful for search.

\subsection{Examples}
In Figures~\ref{algo_in_repo}, \ref{alg3}, and \ref{algo_in_github}, we show a few examples of algorithm implementations found from the {\it Stony Brook code database} and {\it GitHub code database}. The implementations of the {\it floyd-warshall}, {\it lcs-length}, and {\it insertion sort} algorithms are the first ranked in \toolname output list. The figures show that pseudo code and implementations are comparable but not exactly the same. For example, in Figure~\ref{code1}, lines~2--5 initialize the matrix {\tt sums} ({\tt D} at line~3 in Figure~\ref{pc1}), and lines~7---11 implement lines 4--7 in Figure~\ref{pc1}. When manually confirming these code fragments, we found that the control flow and math operators are important here. The figures also show that mapping the code and natural language to the same vector space is important and helpful for search. For example, pseudo code in Figure~\ref{fig:example_repo_pc1}  initializes matrices using natural language at line~4, and in Figure~\ref{fig:example_repo_code1}, source code at lines~3---4 implement line~4 in Figure~\ref{fig:example_repo_pc1}. Pseudo code in Figure~\ref{pc} uses ``$<$'' at line~5, and source code in  Figure~\ref{code} uses the function ``{\tt less}'' at line~4. Since we map both code and natural language to the same vector space using CodeBERT, \toolname is able to retrieve such relevant code.

\section{Threats to Validity}

\begin{figure}
\tiny
	\begin{subfigure}[t]{0.45\columnwidth}
			\lstset{mathescape = false, numbers = left, basicstyle= \footnotesize, mathescape=true}
\begin{lstlisting}[style=mystyle-java]
FLOYD-WARSHALL(W) 
n = W.rows
D=W
for k = 1 to n
	for i = 1 to n
		for j =1  to n
			d_ij = min(d_ij,d_ik+d_kj)
return D
\end{lstlisting}			\caption{Pseudo code} \label{pc1}
	\end{subfigure}
	\begin{subfigure}[t]{0.9\columnwidth}
		       \lstset{numbers = left, breaklines=true, basicstyle=\footnotesize, mathescape=true} %,escapeinside={<@}{@>}}
\begin{lstlisting}[style = mystyle-java]
...
final int[][] sums = new int[vertices.size()][vertices.size()];
for (int i = 0; i < sums.length; i++) {
    for (int j = 0; j < sums[i].length; j++) {
        sums[i][j] = Integer.MAX_VALUE;}}
...
for (int k = 0; k < vertices.size(); k++) {
    for (int i = 0; i < vertices.size(); i++) {
        for (int j = 0; j < vertices.size(); j++) {
			...
        	final int summed = (ikCost != Integer.MAX_VALUE && kjCost != Integer.MAX_VALUE) ? (ikCost + kjCost) : Integer.MAX_VALUE;
        	...
...
return allShortestPaths;
...
\end{lstlisting}
	    \caption{Real-World Implementation, a segment from a function} \label{code1}		
	\end{subfigure}
	\caption{{\tt Floyd Warshall} in  {\it Stony Brook code database}}~\label{algo_in_repo}
		    \vspace{-0.1cm}

\end{figure}
\begin{figure}
\tiny
	\begin{subfigure}[t]{0.49\columnwidth}
			\lstset{mathescape = false, numbers = left, basicstyle= \footnotesize, mathescape=true}
\begin{lstlisting}[style=mystyle-java]
LCS-LENGTH(X,Y)
m = X.length
n = Y.length
let b[1..m,1..n] and c[0..m,0..n] 
be new tables
for i = 1 to m
	c[i,0] = 0
for j = 0 to n
	c[0,j] = 0
for i = 1 to m
	for j = 1 to n
		if x_i == y_j
			c[i,j] = c[i-1,j-1] + 1
			b[i,j] = "\nwarrow"
		else if c[i-1,j] \geq c[i,j-1]
			c[i,j] = c[i-1,j]
			b[i,j] = "\uparrow"
		else if c[i,j] = c[i,j-1]
			b[i,j] = "\leftarrow"                   
return c and b
\end{lstlisting}
			\caption{Pseudo code} \label{fig:example_repo_pc1}
		
	\end{subfigure}
	\begin{subfigure}[t]{0.48\columnwidth}
		       \lstset{numbers = left, breaklines=true, basicstyle=\footnotesize, mathescape=true}
\begin{lstlisting}[style = mystyle-java]
int n = a.length();
int m = b.length();
int S[][] = new int[n+1][m+1];
int R[][] = new int[n+1][m+1];
int ii, jj;
for(ii = 0; ii <= n; ++ii) {
	S[ii][0] = 0;
	R[ii][0] = UP;}
for(jj = 0; jj <= m; ++jj) {
	S[0][jj] = 0;
	R[0][jj] = LEFT;}
for(ii = 1; ii <= n; ++ii) {
	for(jj = 1; jj <= m; ++jj) { 
		if( a.charAt(ii-1) == b.charAt(jj-1) ) {
			S[ii][jj] = S[ii-1][jj-1] + 1;
			R[ii][jj] = UP_AND_LEFT;}
		else {
			S[ii][jj] = S[ii-1][jj-1] + 0;
			R[ii][jj] = NEITHER;}
		if( S[ii-1][jj] >= S[ii][jj] ) {	
			S[ii][jj] = S[ii-1][jj];
			R[ii][jj] = UP;}
		if( S[ii][jj-1] >= S[ii][jj] ) {
			S[ii][jj] = S[ii][jj-1];
			R[ii][jj] = LEFT;}}}
...
return new String(lcs);
\end{lstlisting}
	    \caption{Real-World Implementation, a segment from a function} \label{fig:example_repo_code1}		
	\end{subfigure}
	\caption{{\tt LCS length} in {\it Stony Brook Algorithm Repository}~\label{alg3}}
\end{figure}
\begin{figure}
\tiny
	\begin{subfigure}[t]{0.4\columnwidth}
			\lstset{mathescape = false, numbers = left, basicstyle= \footnotesize, mathescape=true}
\begin{lstlisting}[style=mystyle-java]
InsertionSort(A)
A[0] = -\infinity
for i = 2 to n
	j = i
	while A[j] < A[j - 1]
		swap(A[j],A[j-1])
		j = j - 1
\end{lstlisting}
			\caption{Pseudo code} \label{pc}
		
	\end{subfigure}
	\begin{subfigure}[t]{0.58\columnwidth}
		       \lstset{numbers = left, breaklines=true, basicstyle=\footnotesize, mathescape=true}
\begin{lstlisting}[style = mystyle-java]
for (int i = 1; i < array.length; i++) {
    T insertValue = array[i];
    int j;
    for (j = i - 1; j >= 0 && less(insertValue, array[j]); j--) {
        array[j + 1] = array[j];
    }
    if (j != i - 1) {
        array[j + 1] = insertValue;
    }
}
return array;
\end{lstlisting}
	    \caption{Real-World Implementation, an entire function} \label{code}		
	\end{subfigure}
	\caption{{\tt Insertion sort} in {\it GitHub code database}}~\label{algo_in_github}

\end{figure}

{\it External threats}: We have used the algorithms from two textbooks in our evaluation. It covers a variety of algorithms; some have more natural language descriptions, some have more math expressions, and some have complex control flow. We have used a variety of C and Java open-source projects from the {\it Stony Brook Algorithm Repository} and popular projects from {\it GitHub}. However, our conclusions may not be generalized to all algorithms and software repositories available.

{\it Internal threats}: We have used the ground truth documented in the  {\it Stony Brook Algorithm Repository} to confirm the correctness of our results. We also used two authors who followed a code review protocol to inspect the search results. Although we have made our best efforts to confirm the results, we may make mistakes, especially for the experiments involving {\it GitHub} repositories where we do not have labeled ground truth. When selecting baselines, we have not found any tools that search for algorithm implementations based on pseudo code. So we used the best-matched SOTA neural search tools, DCS and OCoR. Our datasets and queries may not be the best fit for their models. Finally, F-rank may be a metric that is biased toward the baselines. The baseline methods that report a function as the search result can have a better F-rank score than \toolname, because the number of candidate code fragments can be significantly bigger than the number of methods in the source code, especially when the project is very small. But code fragment is often a much more precise output for real-world code.  
\section{RELATED WORK}
With recent advancements in deep learning, extensive work has been performed on code search, namely {\it neural search}. We have categorized the related works into three sections; the first two summarize the code search techniques based on different input queries, and the last is about code representation learning.

\vspace{0.3cm}
\noindent{\bf Code search using natural language as a query.} This line of work aims to enable users to search source code with natural language, where the output is code snippets ranked by their similarities to the query. Earlier work has designed ranking functions with engineered features~\cite{Lv2015,Mcmillan2013}. More recently, the task has been modeled as a cross-modality search problem since the query (natural language), and the target (source code) are of different types. The state-of-the-art methods apply deep learning techniques to encode both modalities into a common semantic vector space where matching pairs are mapped close to each other~\cite{haldar2020multi,ling2021deep,gu2018deep,zhu2020ocor,bui2021infercode}. \cite{gu2018deep} is one of the first works in this line. It applies LSTM and Multi-Layer Perceptron to encode both modalities, where both code snippets and natural language descriptions are considered sequences. There are several following works~\cite{cambronero2019deep,husain2019codesearchnet,feng2020codebert} exploiting different encoding models. 
Later works further consider the underlying structure of source code to obtain better representations, such as presenting the source code by Abstract Syntax Trees (ASTs) or AST-based graphs \cite{haldar2020multi,ling2021deep, 2021iclrgraphcodebert} and CFGs \cite{wan2019multi}.

\vspace{0.2cm}
\noindent{\bf Code search using code snippet as a query.} Here, the task is to find code clones and code that has similar functionalities. Lexical and syntactic features are often used to compare the similarity of two code snippets~\cite{white2016deep,yang2015classification,li2017cclearner}. FaCoy~\cite{kim2018facoy} uses query alternation to enrich the query to find code fragments with similar functionality. {\it DeepSim}~\cite{deepsim} extracts the features from the program's data flow and control flow for detecting code clones and function similarities. SLACC~\cite{slacc} uses dynamic analysis and can find similar functionalities across Python and Java.

This paper focuses on the task of code search by pseudo code. Pseudo code uses a mixture of natural language descriptions and mathematical expressions, similar to a code snippet. Compared to natural language queries, pseudo code is more structured since it describes the control flow of the computation steps. Compared to code query, it is language-independent and contains fewer irrelevant implementation-level details. To the best of our knowledge, we are the first to enable automatic analysis of pseudo code. SPoC~\cite{kulal2019spoc} is the first work that generates code based on pseudo code. \cite{ase15} is the first work that generates pseudo code from code.

\vspace{0.3cm}
{\bf Code representation learning.} The success of representation learning techniques in NLP inspired many works on code representation learning. Following  word2vec~\cite{mikolov2013efficient}, {\it Code2vec}~\cite{alon2019code2vec} and {\it Code2seq}~\cite{alon2018code2seq} encode a sequence of tokens along the paths traversing AST as code embedding. The models are useful for predicting method names and performing code summarization. Inspired by pre-trained transformer-based language model BERT~\cite{devlin2018bert}, CodeBERT~\cite{feng2020codebert} and GraphCodeBERT~\cite{2021iclrgraphcodebert} also use transformers for pretraining code representations. We used pre-trained CodeBERT to capture the natural and programming language information in pseudo code and source code. {\it Flow2Vec}~\cite{flow2vec} leverages the interprocedural value flow to generate code embedding and can perform code classification and summarization. {\it LiGer}~\cite{blend} uses a selection of blended symbolic and concrete traces, and DYPRO~\cite{dynamic}  uses dynamic traces for code embedding, both of which use RNN and are shown to be useful for code classification.

\section{Conclusions}
This paper presents a framework \toolname that can establish the mappings between pseudo code and source code and search real-world open source projects for algorithm implementations. \toolname consists of three components: (1) automatic algorithm analysis, (2) representation learning for source code and pseudo code, and (3) vector-based code retrieval. It compared the algorithms and source code using control flow, math, and natural language. \toolname can search for both C and Java code. It found a total of $74$ algorithm implementations out of $87$ queries in the {\it GitHub} projects and located $64$ implementations given $65$ algorithm queries. In the {\it Stony Brook Algorithm Repository}, with the majority ($>$90\%) of the correct implementations ranked in the top 10 search results. In the future, we plan to experiment with more types of algorithms and also study the pairs of pseudo code and source code for applications.
\bibliographystyle{ACM-Reference-Format}
\bibliography{references}
\end{document}